\documentclass{article}
\usepackage{graphicx}%
\usepackage{multirow}%
\usepackage{array}
\usepackage{amsmath,amssymb,amsfonts}%
\usepackage{amsthm}
\usepackage[title]{appendix}%
\usepackage{xcolor}%
\usepackage{textcomp}%
\usepackage{manyfoot}%
\usepackage{algorithm}%
\usepackage{algorithmicx}%
\usepackage{algpseudocode}%
\usepackage{fix-cm}
\usepackage{float}
\usepackage[utf8]{inputenc}
\usepackage[numbers,sort&compress]{natbib}
\usepackage{listings}%
\usepackage{hyperref}
\usepackage{tabularx}
\usepackage{enumerate}
\usepackage{braket}
\usepackage{dirtytalk}
\usepackage{siunitx}
\usepackage{caption}
\usepackage{booktabs}
\usepackage{geometry}

\usepackage{subfig}
\usepackage{mathtools}
\DeclareUnicodeCharacter{039B}{\ensuremath{\Lambda}}
\DeclareMathSizes{8.43146}{8.43146}{6.8}{6.8}

\usepackage{urwchancal}

\begin{document}

\title{Quintessence and False vacuum: Two sides of the same coin?}

\author{
  Pradosh Keshav MV\textsuperscript{1} \and
  Kenath Arun\textsuperscript{2}\thanks{Email: kenath.arun@christuniversity.in}
}

\date{
  \textsuperscript{1,2}Department of Physics and Electronics, Christ University, Bangalore 560029, Karnataka, India
}

\maketitle

\begin{abstract}
    We study late-time acceleration scenarios using a quintessence field initially trapped in a metastable false vacuum state. The false vacuum has non-zero vacuum energy and could drive exponential expansion if not coupled with gravity. Upon decay of the false vacuum, the quintessence field is released and begins to evolve. We assumed conditions where the effective scalar potential gradient must satisfy \(\nabla V_{\text{eff}} > A\), characterized by a pressure term approximately \(\Delta p / p > \mathcal {O} (\hbar)\) invoking the recently proposed string swampland criteria. We then derived the effective potential of the scalar with an upper bound on the coupling constant \(\lambda < 0.6\). Further analysis revealed that \(V_{\text{eff}}\) shows a slow-roll behavior for \(0.1 > \lambda > -0.04\) in the effective dark energy equation of state (EoS) \(-0.8 < w_0 < -0.4\), stabilizing at points between \(1 < A < 2.718\). Our results suggest a stable scalar decoupled from its initial meta-stable state, could indeed lead to a more stable universe at late times. However, slight deviations in parameter orders could potentially violate the swampland criteria if \(V_{\text{eff}}\) grows too rapidly. Since this is not something we expect, it opens up the possibility that the current dark energy configuration might be a result of a slowly varying scalar rather than being arbitrary. 
\end{abstract}

\newpage
\section{Introduction}
The accelerated expansion of the universe is attributed to dark energy \citep{perlmutter1999measurements, riess1998observational}. This exotic fluid, characterized by an equation of state (EoS) of \(w=-1\), forms the basis of \(\Lambda\)-CDM model alongside Cold Dark Matter (CDM). Despite its agreement with most observations, the \(\Lambda\)-CDM model faces significant challenges that necessities further investigation  \citep{delpopolo2016, turner2018, weinberg2001}. One major issue is the old cosmological constant problem \citep{lombriser2019cosmological, padilla2015lectures, ng1992cosmological, barrow2011new, garriga2001solutions}. According to quantum field theory, the vacuum should also have energy density due to virtual particles and fluctuations. This energy should contribute to the cosmological constant. But observations show that it has an energy scale of approximately \((10^{-3} \, \text{eV})^4\) on the order $\sim 10^{120}$ smaller than expected\citep{carroll2001cosmological}. Another issue arises regarding the small but non-zero value of the cosmological constant and why it has the specific value that it does in the present-day universe. This is known as the new cosmological constant problem. The new problem also motivates the \say{cosmic coincidence problem,} asking why the energy density of the cosmological constant (dark energy) is of the same order of magnitude as the matter density in the current epoch of the universe. 

A possible resolution in \citep{padmanabhan2003cosmological} considers an effective cosmological constant $\Lambda_{\text{eff}}$ with two effects occurring simultaneously. One is the intrinsic contribution from the gravitational Lagrangian of form $R-2\Lambda$, where $R$ is the Ricci scalar, and contributes to the spacetime curvature even in a vacuum (when the energy-momentum tensor $T_{ik}=0$ ). Another is the dynamic shifts in the matter Lagrangian, where evolving energy densities, such as those from a scalar field $\phi$, can modify the cosmological constant over time. For example, if a scalar field \(\phi\) settles into a minimum of its potential \(V(\phi)\), this minimum value \(V(\phi_{\text{min}})\) contributes a constant term to the energy-momentum tensor. This effectively resembles a cosmological constant. But such scenarios are later found conflicting as it requires $\Lambda$ and \(V(\phi_{\text{min}})\) to be extremely fine-tuned for the bound $|\Lambda_{\text{eff}}|< 10^{-56} \text{cm}^{-2}$ to be satisfied \citep{sahni2002cosmological}. This could change \(V(\phi_{\text{min}})\) by several orders of magnitude during the evolution of the universe.

The latter case of $\Lambda_{\text{eff}}$ could naturally lead one to ask whether the universe could currently be in a false vacuum state - a metastable state that might eventually decay into a true vacuum, potentially leading to a vanishing cosmological constant. The false vacuum is a temporary state that can transition to a lower-energy true vacuum through a process known as vacuum decay. Though it is widely studied in \citep{ barr2001cosmological, bousso2015inflation, blau1987dynamics, krauss2008late, rafelski2015dynamical, fischler1990quantization, bronnikov2001spherically}, applying false vacuum scenarios to phenomenological dark energy models is challenging due to large fluctuations triggering vacuum decay in the early universe \citep{kost2022massless, smeenk2005false, sato1981first}. This extends to any dynamic model of false vacuum, which may face challenges in explaining various symmetry-breaking phenomena, such as those observed in electroweak interactions and quantum chromodynamics (QCD). Hence, most phenomenological dark energy models often conflict with false vacuum scenarios as any transient phenomenon or transition associated with non-zero \(\Lambda\) should be fundamental \citep{mersini2006}.

Adding to the complexity are the swampland criteria, which have been proposed as fundamental constraints on effective field theories derived from string theory and, by extension, on any consistent theory of quantum gravity \citep{eichhorn2024absolute, palti2019swampland}. These criteria impose limits on the potential \say{$V$} of scalar fields, such as requiring steep potentials for field excursions, which essentially limit the notion of stable de Sitter vacua. Some models attempt to avoid these constraints by considering $V(\phi)$ with local minima at positive values, leading to a stable de Sitter vacua at late times. The potential may be positive, but the scalar field might not be at a minimum, as seen in most quintessence models \citep{obied2018}. However, the stability remains plausible as long as the magnitude of the gradient $|\nabla V|$ remains sufficiently small and comparable to the potential $V$ itself. Tighter constraints on $|\nabla V| > A$ for a constant $A > 0$ can lead to field excursions, but the existence of supersymmetric vacua needs gravity to stabilize vacuum \citep{cosemans2005stable, espinosa2020stabilizing, cvetivc1993non}. To avoid the latter, one can assume a scenario where $A$ is a function of $\phi$ such that $A(\phi) \leq 0$. This means that $V(\phi)$, normalized to the order of the parameter $A$, the derivative of $\phi$ shouldn't be lesser than the lower bound of the quintessence potential. However, the stability of such configurations remains a significant concern while taking total contribution to potential, including the Higgs.

In light of these theoretical challenges, it is useful to revisit Gliner's early work on cosmological models that feature a constant vacuum energy density $\rho_{\text{vac}}$ leading to exponential expansion and a de Sitter space  \citep{gliner1965, gliner1970, gliner1975}. Gliner's model shows how a constant vacuum energy density can lead to the exponential expansion of the universe, resulting in a de Sitter space. This scenario is characterized by the scale factor $a(t) \propto \exp(\chi t)$ where $\chi = \frac{8 \pi G}{3}\rho_{\text{vac}}$. During the rapid expansion, $\rho_{\text{vac}}$ remains constant, while other energy densities, such as matter and radiation, experience quick dilution. This idea was applied successfully by Starobinsky and Guth to early inflation scenarios \citep{starobinsky1979, guth1981}. Another significant contribution is from Wald \citep{wald1983}, who noted that FLRW solutions to the Einstein equation with a positive $\Lambda$ generally approach the exponential vacuum solution with a Hubble rate $\sqrt{\Lambda/3}$, provided that the stress-energy tensor satisfies specific energy conditions in terms of two inequalities. This implies solutions to the semiclassical Einstein equations align well with classical vacuum solutions for a positive cosmological constant $\Lambda$. The de Sitter solution with constant curvature were also considered in \citep{juarez-aubry2019}.

However, Wald's work could be traced back to Coleman's early works on vacuum instability \citep{coleman1977, coleman1980, callan1977}. He was one of the first to highlight scenarios where a physical system does not reside in an absolute energy minimum but instead exists in a state separated by an effective potential barrier. Additionally, \citep{wetterich1988cosmology, wetterich2004phenomenological, stachowski2016} studied some interpretations of dark energy connected to the transition framed as a consequence of a time-dependent cosmological constant, where the vacuum state is determined by the dynamics of underlying semi-classical theory. This is later reflected in various degrees of the bounce action related to the disparity in pressure corrections between the true and false vacuum states. A similar state has been recently considered for Schwarzschild black holes with no quantum corrections up to second order in curvature \citep{calmet2017}. The possibility of using non-minimal scalar fields without affecting the Higgs mass is detailed in \citep{masina2012, masina2012b}. The generalities of string landscape and effective field theories studied by \citep{brennan2017, vanBeest2022, agmon2022, heisenberg2018} further necessitate fine-tuning into the stability problem.

To investigate this further, we take a semi-classical approach to classical scalar fields and set a pressure bound to the effective potential as $\Delta p/p_f > \mathcal{O}(\hbar)$, where $p_f$ is the pressure related to a false vacuum bubble during decay. Similarly, \citep{hertzberg2019, franca2002} has studied simple string compactifications of four-manifolds when both $V$ and $\nabla V$ are fine-tuned since there is only a single fine-tuning required using the relation $|\nabla V| \sim \frac{V}{M_{pl}}$. However, we do not expect the same; rather, with a lower bound on the coupling constant $\lambda > 0.1$ (against the expected $\lambda < 0.6$ \citep{han2019}), we follow \citep{axenides2004} with a slowly evolving effective potential $V_{\text{eff}}$ between $-0.04 <\lambda <0.1$, when $\nabla V_{\text{eff}} > A$ is maximum. This is similar to the scalar potential usually employed in Higgs' instability scenarios during inflation-like conditions when the growth and potential of the false vacuum bubble are stable \citep{agrawal2018}. 

The paper is organized as follows. We begin by analyzing the metastable false vacuum state and its stability in Sections 2 and 3. In Section 4, we derive the bounce action associated with vacuum decay, focusing on semi-classical conditions for a reduced bounce action. Section 5 discusses the effective potential \(V_{\text{eff}}\) and its role in maintaining vacuum stability, particularly in the context of the string landscape. We also examine the influence of dark energy density \(\rho_{de}\) on cosmic dynamics and the stability of the potential, which is further explored in Sections 6 and 7.

\section{Pressure Correction to Energy-Momentum Tensor}
In transitioning from a metastable false vacuum to a true vacuum, pressure corrections to the energy-momentum tensor significantly influence the scalar field dynamics \citep{Lee2006, Batini2024, copeland1994false}. The first quantum corrections to the leading semiclassical approximation of the decay rate of an unstable ground state are calculated in \citep{callan1977}.  To classically analyse the pressure difference between the false and true vacuum states, we assume a bound \(\frac{\Delta p}{p} \geq \mathcal{O}(\hbar)\), which analytically corresponds to the ratio of kinetic to the potential energy of the scalar. 

The energy-momentum tensor for a scalar field \(\phi\) is given by:
\begin{equation}
    T_{\mu \nu} = \partial_{\mu} \phi \partial_{\nu} \phi - g_{\mu \nu} \left(\frac{1}{2} \partial^{\alpha} \phi \partial_{\alpha} \phi + V(\phi)\right).
\end{equation}where \(V(\phi)\) represents the potential governing the scalar field. We distinguish between the false vacuum potential \(V_f(\phi)\) and the true vacuum potential \(V_t(\phi)\). For the false vacuum, the potential \(V_f(\phi)\) can be expressed as:
\begin{equation}
    V_f(\phi) = \frac{1}{2} m_{f}^2 (\phi - \phi_f)^2.
\end{equation} where \(m_f\) is the mass parameter associated with the scalar field in the false vacuum state. The true vacuum potential \(V_t(\phi)\), on the other hand, is treated as a constant, representing the lowest energy state.

The pressure difference $\Delta p$ as the scalar field $\phi$ transitions from the false to the true vacuum can be calculated by focusing on the time component of the energy-momentum tensor $T_{00}$:
\begin{equation}
    \Delta p = p_t - p_f = -\frac{1}{2} \left(T_{00, t} - T_{00, f}\right)
\end{equation}where \(p_f\) and \(p_t\) are the pressures in the false and true vacuums, respectively. The corresponding density and pressure terms are computed as follows:
\begin{equation}
    \rho_f = \frac{1}{2} (\partial_0 \phi_f)^2 + V_f(\phi_f)
\end{equation}
\begin{equation}
    p_f = \frac{1}{2} (\partial_0 \phi_f)^2 - V_f(\phi_f).
\end{equation}For the true vacuum state $\phi = \phi_t$, assuming that $V_t(\phi)$ is at energy minimum, the kinetic term vanishes, yielding:
\begin{equation}
    \rho_t = V_t(\phi_t)
\end{equation}
\begin{equation}
    p_t = -V_t(\phi_t)
\end{equation}With the above expressions, we can compute the pressure difference $\Delta p$ as follows:
\begin{equation}
    \Delta p = \left(-V_t(\phi_t)\right) - \left(\frac{1}{2} (\partial_0 \phi_f)^2 - V_f(\phi_f)\right)
\end{equation}

Quantum corrections also play a significant role in modifying the pressure difference during the vacuum transition. These corrections can be expressed through a perturbative expansion:
\begin{equation}
    \Delta p = \Delta p^{(0)} + \Delta p^{(1)} + \Delta p^{(2)} + \dots,
\end{equation}
where \(\Delta p^{(0)}\) represents the leading term, which is essentially the bounce action at the zeroth order $B^{(0)}$ corresponding to the classical energy difference between the false and true vacuums, and \(\Delta p^{(1)}\), \(\Delta p^{(2)}\), etc., represent higher-order quantum corrections.

\begin{figure}[H]
    \centering
    \includegraphics[width=0.7\linewidth]{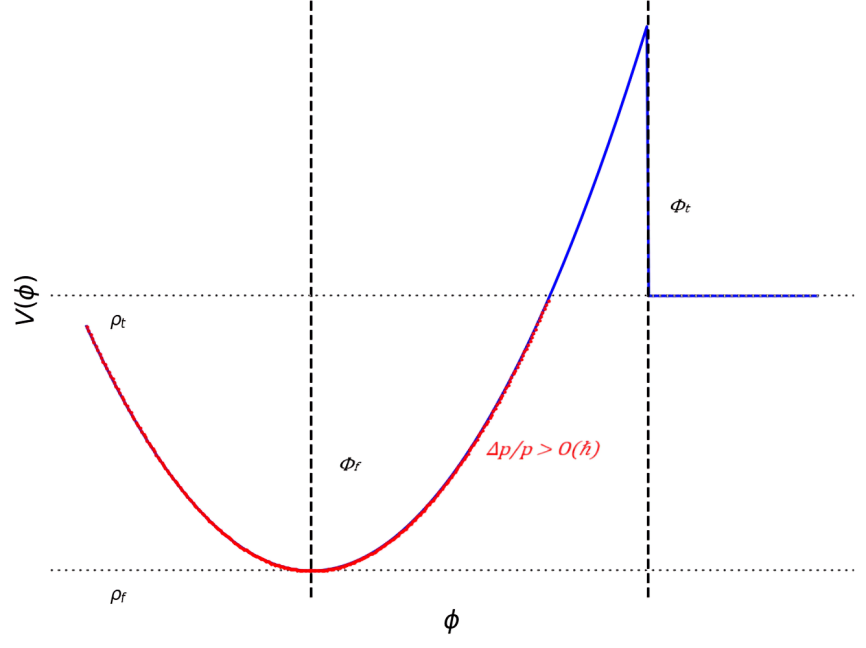}
    \caption{\small Evolution of scalar field between true and false vacuum: The red curve represents the potential $V(\phi)$ in the false vacuum, and the field oscillates until it tunnels to the true vacuum represented by the blue curve.}
    \label{fig:falsevaccum}
\end{figure}

The leading-order correction to the pressure ratio, \(\frac{\Delta p}{p_f}\), can be approximated by:
\begin{equation}
    \frac{\Delta p}{p_f} \approx 1 - \frac{1}{2} \frac{(\partial_0 \phi_f)^2}{p_f}
\end{equation}
indicating that the potential energy difference primarily drives the pressure difference (refer to \citep{calmet2018}), with a smaller contribution from the kinetic term as depicted in Fig. \ref{fig:falsevaccum}. 

To further simplify the analysis, we assume a Yukawa-type interaction where the scalar field \(\phi\) decouples from all other fields except those in the false vacuum state \citep{kowalska2022naturally,gies2017impact}. Suppose the kinetic term dominates over $p_f$ at early times and $\hbar$ has a small value, we consider $\frac{\Delta p^{(0)}}{p_{\text{(today)}}} > \mathcal{O}(\hbar)$ holds true even at late times. i.e., for \(\partial_0 \phi_f \sim M_p\) (Planck mass, \(1.22 \times 10^{19}\) GeV) and \(p_f = (10^{16} \text{ GeV})^4\). This assumption holds when the mass of the scalar field \(M_0 > 10^{-13}\) eV, leading to the following relationship:
\begin{equation}
    \frac{(\partial_0 \phi_f)^2}{p_f} \approx \frac{(M_p)^2}{(10^{16} \text{ GeV})^4} \sim 10^{-26}\text{ GeV}^{-2},
\end{equation}
where \(M_p\) is the Planck mass. This small ratio implies that if the Planck mass (or $M_0$ close to it) is much larger than the energy scales involved in the vacuum transition, the kinetic-to-potential ratio will be very small, indicating a more stable state at late times.

Following equation (11), equation (10) reduces to \(\frac{\Delta p}{p_f} \approx 1\) when the classical contribution \(\Delta  p^{(0)}\), resulting from the transition is comparable to the pressure \(p_f\) in the initial state. This means the contribution of the kinetic energy term to \(\Delta p\) is significant but not so large as to dominate over the potential energy difference.

Considering a small change in $\Delta p^{(0)}$ around weak-scale coupling at strong energy conditions, we expect quantum corrections of order ($\Delta p^{(1)} + \Delta p^{(2)} + \dots$). This implies $<T_{ab}^{rev}>_w \geq H_0$, which is the vacuum expectation value of the renormalized stress-energy tensor. This contribution is significant at high energy scales such as the Higgs, where the values below contain quantum fluctuations on the order of $10^{13}$ GeV\citep{kohri2016higgs}.
\section{False Vacuum and Running Cosmological constant }
Recent studies, particularly those by Urbanowski \citep{urbanowski2022, urbanowski2023}, have investigated metastable false vacuum, proposing that the energy of such a state reflects the characteristics of a running cosmological constant, $\Lambda(t)$. In his framework, the energy $E(t)$ of a quantum system within the metastable state is examined through various decay phases. The false vacuum, representing a local energy minimum, is contrasted with the true vacuum, which corresponds to the lowest energy state. The stability of a field trapped in the false vacuum can be maintained if the energy difference \(E_0 - E_R\) exceeds the bare cosmological constant. This suggests a possible equivalence between the false vacuum energy density \(\rho_f\) and the dark energy density \(\rho_{de}\), or varying potential with time-dependent cosmological constant \(\Lambda(t)\).

The decay rate of the false vacuum, \(\Gamma\), is expressed as:
\begin{equation}
    \Gamma \propto \exp\left(-\frac{B}{\hbar} + \Lambda(t)\right) [1+\mathcal{O}(\hbar)]
\end{equation}
where \(\Lambda(t)\) acts as a source term influencing the decay rate, and \(B\) is the bounce action, quantifying the tunnelling process from the false vacuum to the true vacuum. A larger bounce action corresponds to a lower decay rate with \(\mathcal{O}(\hbar)\) term accounting for quantum corrections beyond the leading order. Note that Eq. (12) is also consistent with Quantum Field Theory (QFT) in curved spacetime \citep{hollands2015quantum, kay1992quantum}, where the time-dependent vacuum energy density \(\rho_\Lambda(t)\) is linked to cosmological parameters such as the Hubble rate \(H(t)\) and the scale factor \(a(t)\).

In scenarios involving a flat universe minimally coupled to gravity, the probability of false vacuum decay has been studied by  Kiselev within the context of (1+1) scalar field theory \citep{kiselev1984}. The decay rate per unit volume, \(\Gamma/\mathcal{V}\), is found to be exponentially suppressed for large energy barriers:
\begin{equation}
    \frac{\Gamma}{\mathcal{V}} = \frac{\epsilon^{\alpha}}{2 \pi} \exp\left[-B + \frac{B}{\pi a^2} \left(\left(\frac{3}{2}\right)^3 - \frac{\pi \sqrt{3}}{4}\right)\right],
\end{equation}
calculated for errors in the exponential factor of order $0 < \alpha < 1$. 

These two descriptions of the decay process—one involving a running cosmological constant \(\Lambda(t)\) and the other relying on classical scalar field theory — must ultimately describe the same physical phenomenon. Otherwise, there will be profound consequences, ranging from potential alterations in fundamental forces to shifts in cosmological parameters. See \citep{devoto2022} for a detailed review. 

Since we focus largely on classical physics, we don't discuss the quantum processes in this context. Rater, we approximate a self-coupled scalar to receive quantum corrections from itself. This requires a bound within the order of magnitude of the Hubble parameter \(H \sim 10^{13}\) GeV and a mass scale parameter \(M_0 \geq M_\mathcal{H}\) where \(M_{\mathcal{H}}\) is the Higgs mass. A similar scenario is studied in \citep{degrassi2012} for inflation-like conditions when a scalar confined in a false vacuum yields exponential expansion. Note that we do not test this assertion under vacuum fluctuations using a quintessence scalar non-minimally coupled to gravity, which may trigger quantum tunnelling out of a false vacuum (a topic not discussed within the context of this paper).

\section{Quintessence and Semi-Classical Approximations}

In cosmology, the quintessence is a popular model introduced as an alternative to the cosmological constant, where a dynamical scalar field with a specific potential drives the accelerated expansion of the universe. Peebles and Ratra proposed the quintessence model to address the dark energy problem \citep{peebles2003cosmological, armendariz2000dynamical}. The potential for the quintessence field is often given by:
\begin{equation}
V(\phi) \propto M^4\left[\exp\left(\frac{M}{\phi}\right)-1\right]
\end{equation}
where \(M\) is a free parameter that sets the energy scale of the potential.

In this context, we consider a scenario where the quintessence field, \(\phi(t)\), evolves over time. The dynamics of the field are governed by the equation \citep{ Liddle:2000cg}:
\begin{equation}
\ddot{\phi} + 3H\dot{\phi} + \frac{dV(\phi)}{d\phi} = 0
\end{equation}
where \(H\) is the Hubble parameter, and the field evolves under the influence of the potential \(V(\phi)\). This equation is analogous to a slow-roll inflationary scenario \citep{urbanowski2017properties, sivaram2020a, myrzakulov2015inflation}, where the damping term \(3H\dot{\phi}\) dominates, leading to a gradual decrease in \(\phi(t)\).

For the quintessence model, the effective potential \(V_{\text{eff}}(\phi)\) can also be influenced by additional quantum corrections. To incorporate these effects, we introduce the bounce solution, \(\phi_B(t)\), which characterizes the field's transition between local and global minima:
\begin{equation}
\phi_B(t) = \phi_0 \exp\left(-\int_{t_0}^{t} \lambda \, dt\right)
\end{equation}
where \(\lambda\) is a coupling constant that modulates the decay rate of the field, and \(\phi_0\) is the field's initial value.

The ultraviolet divergences when calculating the quantum corrections to the Euclidean action necessitate the standard renormalization procedure \citep{devoto2022}. Specifically, the classical action needs to be expressed in terms of $B_0$ and supplemented by appropriate counterterms. This modifies the Euclidean action as:
\begin{equation}
    S(\phi) \xrightarrow{} S(\phi) + \hbar S_{ct}(\phi)
\end{equation} where $S_{ct}$ is one loop counter terms. The effective action for the scalar field can then be expressed as:
\begin{equation}
S(\phi) = B_0 - \hbar \int dt \, \lambda \phi(t)
\end{equation}
where \(B_0\) represents the classical contribution to the action, the second term accounts for the quantum corrections.  The effective potential can then be derived as:
\begin{equation}
V_{\text{eff}} = \int_{t_0}^{t} \left[\Lambda(t) + V_0 \exp\left(-\lambda \phi_0(t)\right)\right] dt
\end{equation}
where \(\Lambda(t)\) is a time-dependent function that influences the field's evolution.

Given that the quintessence field behaves similarly to a perfect fluid in this model, the potential \(V(\phi)\) must satisfy certain stability conditions to avoid runaway solutions or the formation of singularities \citep{berezin19883}. This can be examined through the bounce solution, where the field's behaviour near the vacuum state is critical:
\begin{equation}
\frac{\delta B_0}{\delta \phi(t)} = -\hbar \frac{d}{d \phi(t)} \ln\left(\frac{\mathcal{V}}{\exp(\Lambda(t))}\right) - \hbar \lambda = 0
\end{equation} This leads to the bounce solution:
\begin{equation}
    \frac{\delta S(\phi)}{\delta_B \phi}|_{\phi_B = \phi_0} =0
\end{equation}
indicating that \(\phi(t)\) is at a minimum of the effective potential, accounting for both classical and quantum contributions. However, for one-loop order, there is no need for the explicit expression of the correction counter terms where $S(\phi)$ varies drastically between the interval $-0.04 < \lambda <0.1$. We will discuss this in detail in section 7.

To further refine the model, we assume $\mathcal{V} = \frac{\Lambda_0 M_0}{8\pi G} $, with $\Lambda_0 (\sim 10^{56}\text{cm}^{-2})$ and $M_0$ ($>10^{-13}$eV) is a constant that sets the energy scale of the potential. We set the scalar field $\nabla \phi \leq 1$ such that a potential energy
barrier may appear because of the gradient term. This means the effective field spends a large amount of time in the absence of any evolution in $\phi(t)$. Additional contributions independent of the field may appear, accounting for specific energy terms in the system. The boundary configuration for \(\Lambda(t)\) is then:
\begin{equation}
\Lambda(t) = \ln\left(\frac{\Lambda_0 M_0}{8\pi G}\right) - \frac{\lambda}{\hbar} \phi(t) + c_1
\end{equation}
where \(c_1\) is an integration constant.

Following equations (19) and (22), the dynamics of the quintessence field using a semi-classical approximation can be described by the following equations:
\begin{equation}
\langle \phi(t) \rangle = \frac{\langle 0| \phi(t) |0 \rangle}{\langle 0| 0 \rangle},
\end{equation}
\begin{equation}
\langle \phi^2(t) \rangle = \frac{\langle 0| \phi^2(t) |0 \rangle}{\langle 0| 0 \rangle}
\end{equation}
where \(\langle \phi(t) \rangle\) and \(\langle \phi^2(t) \rangle\) represent the functional expectation values of the field and its square, respectively, averaged over the quantum state.

Considering the (time-dependent) influence of the quantum corrections, the effective potential \(V_{\text{eff}}(\phi)\) takes the form:
\begin{equation}
V_{\text{eff}}(\phi) = V(\phi) + \frac{\hbar \lambda}{2} \langle \phi^2(t) \rangle
\end{equation}
This potential accounts for both the classical contribution from \(V(\phi)\) and the quantum corrections involving the expectation value of \(\phi^2(t)\).

The dynamics of the field \(\phi(t)\) can be realized by solving the equation:
\begin{equation}
\ddot{\phi} + 3H\dot{\phi} + \frac{dV_{\text{eff}}(\phi)}{d\phi} = 0
\end{equation} with maximum at:
\begin{equation}
    \frac{dV_{\text{eff}}(\phi)}{d \phi} = 0
\end{equation}
where the quantum corrections are incorporated into the effective potential at its local minimum. In this case, $V_{\text{eff}}$ is an ordinary function of a single canonical scalar that does not contain any derivatives of the field.

Note that some parametrization is required for $V_{\text{eff}}$ to determine whether $\phi(t)$ remains within the order of magnitude constraints. This is because, under semi-classical approximations or inflation-like conditions, the dynamics of the scalar field can be influenced by a source term \citep{senatore2016lectures, struyve2015}. Therefore, considering an exponential potential with a source term is reasonable, as such potentials often appear in unification theories like supergravity and Kaluza-Klein models. However, if one chooses a different form of potential other than the exponential potential, such as power-law or polynomial forms, the bounce solution becomes exponentially suppressed, affecting the decay rate and, consequently, may lead to oscillatory field dynamics or delayed stabilization. While a strong dependency of the decay rate on the scale factor can partially mitigate this suppression at early times, achieving a stable late-time configuration would likely require large field displacements \((\geq M_{{p}})\) as \(\phi\) cannot source accelerated expansion. Mostly because corrections of the form \(\phi^n/M_{p}^{n-4}\) with \(n>4\) may become important, and towers of states typically become light as predicted by the swampland distance conjecture \citep{ooguri2007geometry}. 

\section{Effective Potentials and Swampland Constraints}

In the framework of string theory, the landscape refers to the set of low-energy effective theories that can emerge from string compactifications and are consistent with quantum gravity. However, not all low-energy theories that seem consistent are compatible with a quantum theory of gravity; those that do not form the so-called swampland. The Swampland conjectures impose stringent conditions on the nature of scalar fields and their potentials, particularly constraining dark energy models, such as quintessence.

One of the key Swampland conjectures is the \say{distance conjecture,} which restricts the gradient of the scalar field potential, \( V(\phi) \), by imposing the condition:
\begin{equation}
 |\nabla_\phi V| \geq \frac{c}{M_p} V   
\end{equation}where \( c \) is a constant of order $\mathcal{O}(1)$ and \( M_p \) is the Planck mass. This constraint implies that the scalar field potential must be steep, challenging the flat potentials typically required for slow-roll inflation or quintessence models.

Given these constraints, the effective potential \( V_{\text{eff}}(\phi) \) for the quintessence field must be carefully constructed to satisfy both the Swampland criteria and the observational requirements for late-time cosmic acceleration. Specifically, to accommodate the Swampland conjecture while addressing the cosmological constant problem, one could consider a scenario where the gradient of the potential is small but non-zero, such that:
\begin{equation}
   |\nabla_\phi V_{\text{eff}}|_{\text{today}} \sim 10^{-55} M_p^2 
\end{equation}This gradient is much smaller than what is typically allowed by the Swampland conjecture, necessitating a delicate balancing of the parameters involved. For instance, if the gradient $\frac{| \nabla V_{\text{eff}} |_{\text{today}}}{\Delta V_{\mathcal{H}}} \sim \frac{10^{-120}}{10^{-65}} \sim 10^{-55} M_p^{2}$, it could help to alleviate the tension between the current value of $\Lambda(t)$ and the maximum potential of the Higgs field at $\mathcal{H}=0$ \citep{agrawal2018}.

In semi-classical gravity, the decay of a false vacuum state can occur via quantum tunnelling, leading to the nucleation of true vacuum bubbles. The effective potential plays a crucial role in determining the dynamics of this process. The Euclidean action \( S_E \) governing the tunnelling process is given by:
\begin{equation}
    S_E = \int d^4x \left[ \frac{1}{2} (\nabla \phi)^2 + V(\phi) \right]
\end{equation}where \( \phi \) is the scalar field. With  \((\nabla \phi)^2 <1 \), the decay rate of the false vacuum is exponentially suppressed by the Euclidean action, \( \Gamma \sim \exp(-S_E/\hbar) \).

Symmetric true vacuum bubbles form within the false vacuum background upon nucleation due to quantum fluctuations. The field configuration inside the bubble, \( \phi_B \), evolves as:
\begin{equation}
   \phi_B(\sqrt{|x|^2 + \tau^2}) \rightarrow \phi_B(\sqrt{|x|^2 + t^2}) 
\end{equation}where \( \tau \) is the Euclidean time, and \( t \) is the real time.

To understand the expansion of the bubble in real space, we examine the effective potential \( V_{\text{eff}}(\phi) \) as a function of the scalar field. Assuming $V_{\text{eff}}$ captures the total energy contribution by integrating over a spatial volume and $\phi(t)$ is approximately uniform within the bubble.  The bubble's volume in normalized space $\int d^3 x =1$ can be expressed as:
\begin{equation}
 \mathcal{V}_{\text{Bub}} \approx A \exp(-\lambda \phi(t))   
\end{equation}where \( A = \exp\left(\frac{c_1 - c_2}{\hbar}\right) \) encapsulates the initial conditions of the system and the scale of the bubble volume and \( \lambda \) determines the steepness of the potential. In this context, we set a bound $c_1 - c_2 \geq \hbar$ that the initial conditions are physically reasonable and avoid extremes. When $A$ is in the range 1 to $\exp (1)$, $c_1 - c_2$ are nearly balanced so that the bubble volume is tuned to be near a critical value below which quantum effects are significant.

The total potential \( V_{\text{tot}} \) of the system, including contributions from both the Higgs field \( \mathcal{H} \) and the quintessence field, is given by:
\begin{equation}
  V_{\text{tot}} = V_{\mathcal{H}}(\mathcal{H}) + V_{\text{eff}}(\phi)  
\end{equation}For the system to remain stable, the gradient of the total potential must satisfy the condition:
\begin{equation}
    |\nabla V_{\text{tot}}| \geq A
\end{equation}in the leading semi-classical limit where \begin{equation}
   |\nabla V_{\text{tot}}| = \sqrt{ \left(\lambda_\mathcal{H} (\mathcal{H}^2 - v^2) \mathcal{H}\right)^2 + \left(V (\phi) M^5 \frac{\exp(M/\phi_f)}{\phi_f^2}\right)^2 }
\end{equation} where \( \lambda_{\mathcal{H}} \) is the Higgs self-coupling constant and \( v \) is the vacuum expectation value of the Higgs field. This ensures that the effective potential is sufficiently dominant, contributing to the stability of the field configuration.

To evaluate the stability of the system, we analyze the ratio of the gradient of the total potential to the effective potential. For a typical scenario where \( V_{\text{eff}} \) dominates, we have:
\begin{equation}
    \frac{|\nabla V_{\text{tot}}|}{V_{\text{eff}}} \approx 3.6 \times 10^{36} \text{ GeV}
\end{equation}This indicates a strong dominance of the gradient over the effective potential, ensuring that the field configuration is stable.

However, when the effective potential is parametrized differently, such as \( V_{\text{eff}} \approx \exp(\lambda \phi(t) ) \), and \( \lambda \phi(t) \geq \frac{\Delta p}{p_f} \) is of order unity, the ratio becomes:
\begin{equation}
    \frac{|\nabla V_{\text{tot}}|}{V_{\text{eff}}} \approx 0.222 \text{ GeV}
\end{equation}
In this case, the gradient may not sufficiently dominate, indicating a need for further fine-tuning of the parameters to maintain stability. Table 1 summarizes the fine-tuned parameters necessary to satisfy the condition \( |\nabla V_{\text{tot}}| > A \) for late-time acceleration scenarios, though it requires careful consideration of the potential gradient and the conditions under which vacuum decay occurs. 

\begin{table}[ht]
    \centering
    \caption{Fine-tuned parameters for satisfying \( |\nabla V_{\text{tot}}| > A \).}
    \small 
    \begin{tabular}{|c|c|c|c|}
    \hline
    \textbf{Parameter} & \textbf{Range} & \textbf{Units} & \textbf{Correction Factor} \\
    \hline
    $\mathcal{H}$ & $0.001 \text{ - } 0.004$ & GeV & $10^5$ \\
    \hline
    $\lambda_{\mathcal{H}}$ & $0.001 \text{ - } 0.01$ & dimensionless & $10^2$ \\
    \hline
    $V_0$ & $0.55 \text{ - } 0.84$ & GeV & $10^7$ \\
    \hline
    $M$ & $0 \text{ - } 1$ & GeV & - \\
    \hline
    $\phi_f$ & $1 \text{ - } 2$ & GeV & $10^{18}$ \\
    \hline
    $v$ & $246$ (fixed) & GeV & - \\
    \hline
    \end{tabular}
\end{table}

\section{ Breit-Wigner Distribution  Dynamics}

In the study of quantum systems, the Breit-Wigner (BW) energy distribution is commonly used to describe the probability distribution of energy levels in unstable states, such as particles undergoing decay. This distribution is characterized by a resonance energy \( E_R \) and a decay width, which together define the likelihood of finding a particle at a specific energy level within the resonance state \citep{stachowski2016}.

Dark energy, responsible for the accelerated expansion of the universe, can be described by a time-dependent energy density \(\rho_{de}\). We study a novel parameterization of dark energy density inspired by the Breit-Wigner distribution:
\begin{equation}
    \rho_{de} = \Lambda(t) + E_R \left[1 + \frac{\alpha}{1-\alpha} \mathcal{R} \left(\frac{J(t)}{I(t)}\right) \right],
\end{equation}
where \(\Lambda(t)\) is the cosmological term that evolves over time, \(E_R\) is a residual energy component analogous to the resonance energy, and \(\alpha\) is a dimensionless parameter modulating the contribution of the ratio \(\mathcal{R} \left(\frac{J(t)}{I(t)}\right)\) to the dark energy density. The functions \(J(t)\) and \(I(t)\) represent time-dependent integrals or intervals arising from the dynamics of the underlying scalar field. Refer \citep{szydlowski2017} for the explicit form of intervals.

For a quintessence field \(\phi\), the dark energy density \(\rho_{de}\) is related to the field's potential \(V(\phi)\) and kinetic energy:
\begin{equation}
   \rho_{de} = \frac{1}{2} \dot{\phi}^2 + V_{\text{eff}}(\phi).
\end{equation}
In this context, we consider the potential \(V_{\text{eff}}(\phi)\) (with negligible contribution from kinetic term) as a combination of the time-evolving cosmological term and the Breit-Wigner-inspired contribution:
\begin{equation}
 V_{\text{eff}}(\phi) = \Lambda(t) + E_R \left[1 + \frac{\alpha}{1 - \alpha} \mathcal{R} \left(\frac{J(t)}{I(t)}\right) \right].
\end{equation}

To ensure that our model remains consistent with Swampland constraints, we require that the gradient of the effective potential satisfies a certain steepness criterion:
\begin{equation}
   \left| \frac{dV_{\text{eff}}}{d\phi} \right| > \frac{A}{M_p},
\end{equation}
where \(A\) is a constant related to the potential's steepness and \(M_p\) is the Planck mass. This inequality ensures that the effective potential is steep enough to avoid late-time de Sitter vacua, which are problematic within the Swampland conjecture framework.

To express the gradient of \(V_{\text{eff}}\) with respect to \(\phi\), we consider the time dependence of \(\Lambda(t)\), \(J(t)\), and \(I(t)\), and re-express \(\Lambda(t)\) as \(\Lambda(\phi)\) using:
\begin{equation}
    \frac{d\Lambda(t)}{d\phi} = \frac{d\Lambda(t)}{dt}  \frac{dt}{d\phi} = \dot{\Lambda}(t)  \frac{1}{\dot{\phi}}
\end{equation}
Similarly, for the ratio-dependent term, we have:
\begin{equation}
    \frac{d}{d\phi} \left( \mathcal{R} \left( \frac{J(t)}{I(t)} \right) \right) = \frac{d\mathcal{R}}{d\left(\frac{J(t)}{I(t)}\right)}  \frac{d\left(\frac{J(t)}{I(t)}\right)}{dt}  \frac{dt}{d\phi}
\end{equation}
Combining these, the gradient of \(V_{\text{eff}}\) with respect to \(\phi\) becomes:
\begin{equation}
    \frac{dV_{\text{eff}}}{d\phi} = \frac{d\Lambda(t)}{d\phi} + E_R  \frac{d}{d\phi} \left[ \frac{\alpha}{1 - \alpha} \mathcal{R} \left( \frac{J(t)}{I(t)} \right) \right]
\end{equation} Substituting the derivatives:
\begin{equation}
    \frac{dV_{\text{eff}}}{d\phi} = \dot{\Lambda}(t)  \frac{1}{\dot{\phi}} + E_R \left[ \frac{\alpha}{1 - \alpha}  \frac{d\mathcal{R}}{d\left(\frac{J(t)}{I(t)}\right)}  \frac{d\left(\frac{J(t)}{I(t)}\right)}{dt}  \frac{1}{\dot{\phi}} \right]
\end{equation}We take:
\begin{equation}
    \Lambda'(\phi) = \dot{\Lambda}(t)  \frac{1}{\dot{\phi}} \quad \text{and} \quad \mathcal{R}'(\phi) = \frac{d\mathcal{R}}{d\left(\frac{J(t)}{I(t)}\right)}  \frac{d\left(\frac{J(t)}{I(t)}\right)}{dt}  \frac{1}{\dot{\phi}}
\end{equation}

For practical purposes, the steepness criterion that ensures stability and compliance with Swampland constraints is:
\begin{equation}
    \left| \Lambda'(\phi) + E_R \left[1 + \frac{\alpha}{1 - \alpha} \mathcal{R}'(\phi) \right] \right| > \frac{A}{M_p}
\end{equation}
This condition imposes a lower bound on the potential's gradient, ensuring that the effective potential does not lead to an undesirable late-time de Sitter vacuum. When $E_R$ and $\alpha$ become zero, equation (47) reduces to $\Lambda$-CDM model. Otherwise, the potential is set to avoid scenarios where the dark energy density becomes too high, which could lead to an unstable or excited vacuum state. 

\section{Cosmological implications of Quintessence and False vacuum scenarios  }

The stability of false vacuum with string landscape constraints some quintessence models demanding $c < 0.4$ \citep{heisenberg2018}. We follow equation (47) and substitute $ \Lambda'(\phi) \sim E_0 - E_R $ to the decay rate and compute the effective EoS without further potential ($V_{\text{eff}}$) and gradient calculations. 

\begin{figure}
    \centering
    \includegraphics[width = 0.8\linewidth]{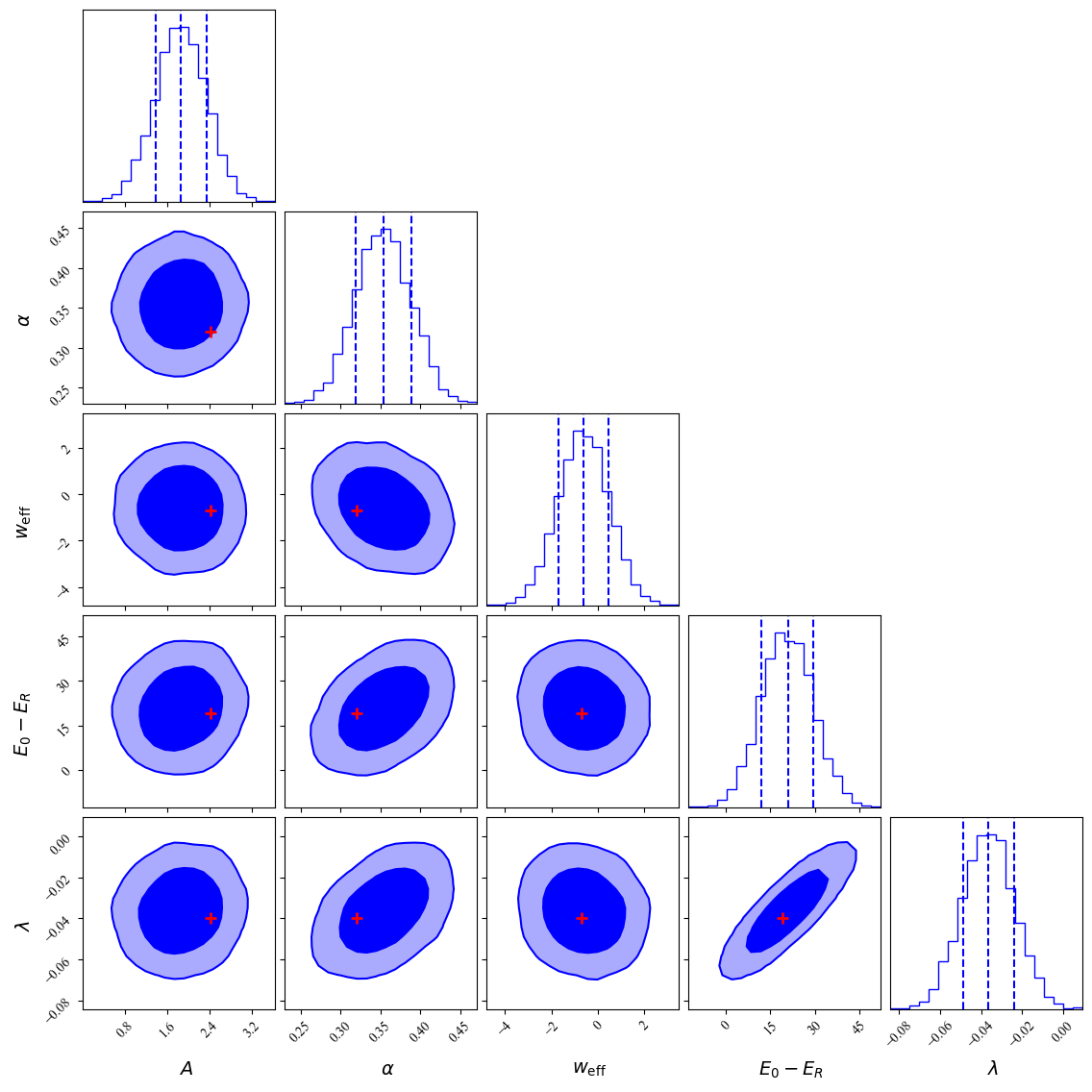}
    \caption{\small Markov Chain Monte Carlo (MCMC) analysis of model parameters following equation (50). The corner plot shows marginalized posterior distributions with contours representing 68\% and 95\% confidence levels, with the maximum likelihood parameters marked in red. The dataset combines Baryon Acoustic Oscillation (BAO), Cosmic Chronometers (CC), and Pantheon+SH0ES Supernova Type Ia data. Prior constraints were set, and likelihood functions were computed using chi-squared statistics for observational data fitting. }
    \label{fig:bestfit1}
\end{figure}

\begin{figure}[H]
    \centering
    \subfloat[$\rho_{de}$ vs $A$]{%
        \includegraphics[width=0.5\textwidth]{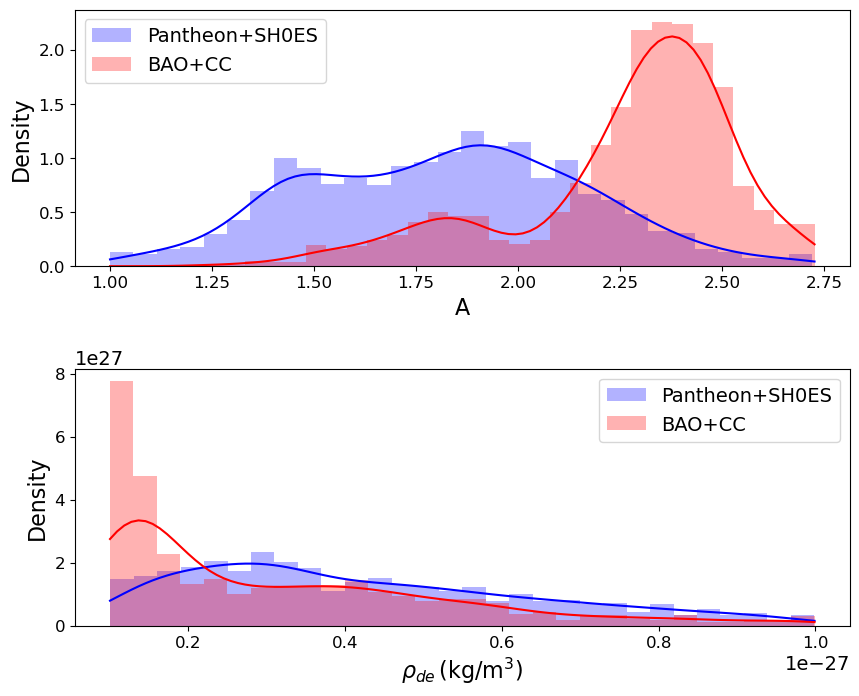}%
        \label{fig:rho_vs_A}%
    }
    \hfill
    \subfloat[$\rho_{de}$ vs $w_{\text{eff}}$]{%
        \includegraphics[width=0.5\textwidth]{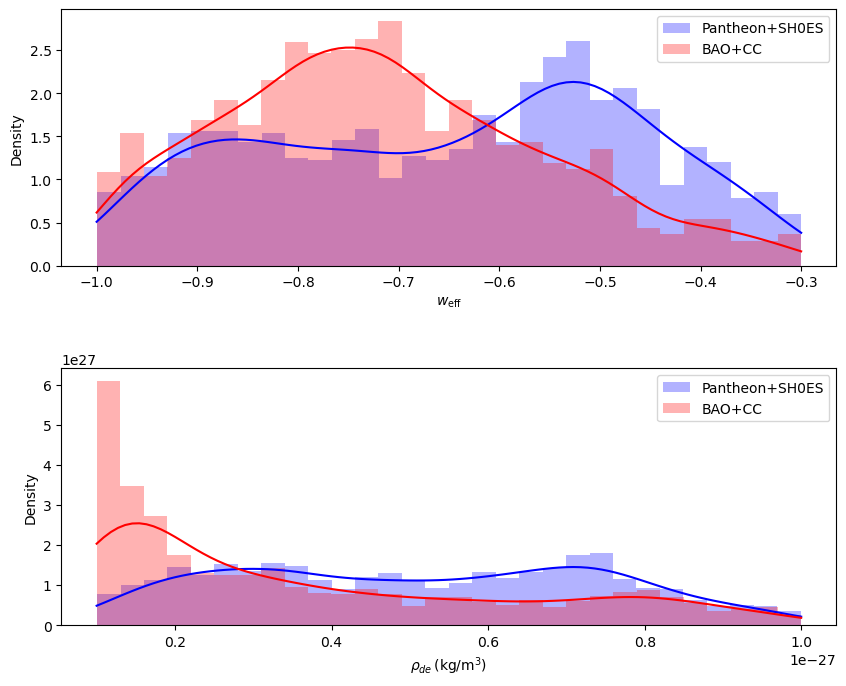}%
        \label{fig:rho_vs_eos}%
    }
    
    \vspace{0.35cm} 
    
    \subfloat[$\rho_{de}$ vs $\lambda$]{%
        \includegraphics[width=0.5\textwidth]{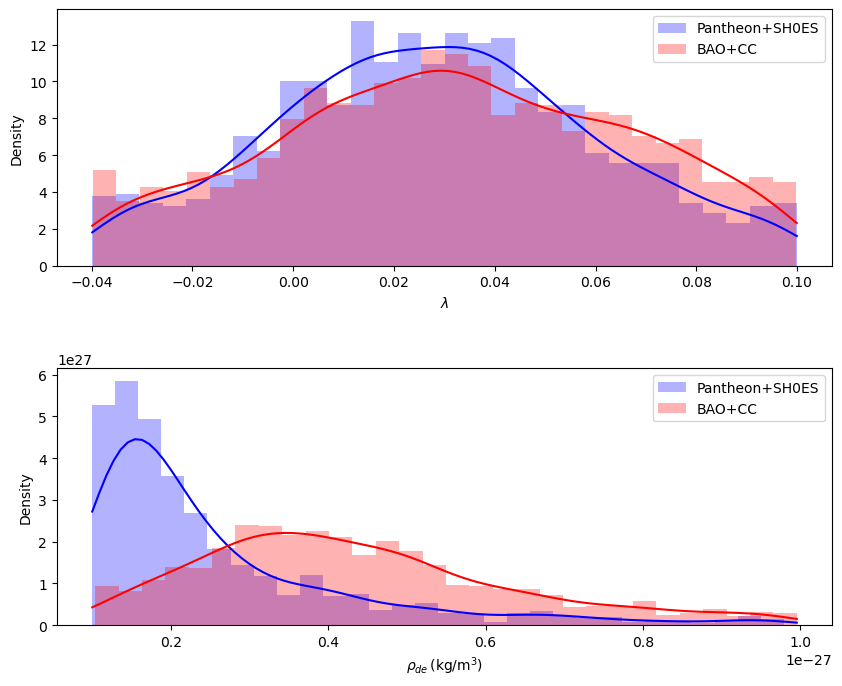}%
        \label{fig:rho_vs_lambda}%
    }
    \hfill
    \subfloat[$\rho_{de}$ vs $\alpha$]{%
        \includegraphics[width=0.5\textwidth]{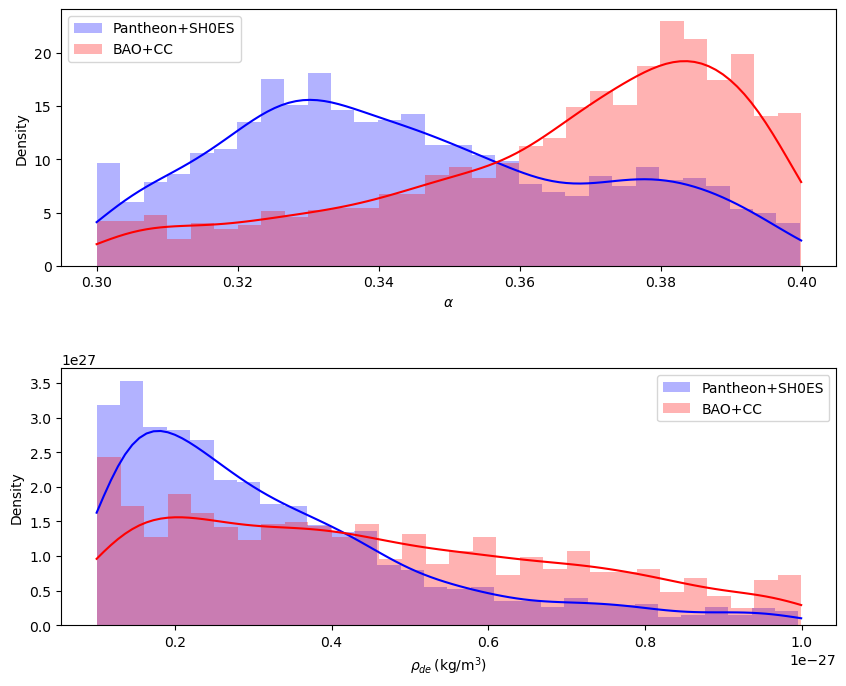}%
        \label{fig:rho_vs_alpha}%
    }
    
    \caption{KDE plot showing the marginalized distributions and pairwise correlations of cosmological parameters obtained from separate MCMC analyses using the Pantheon+SH0ES (blue) and BAO+CC (red) datasets. The histograms along the diagonal represent the posterior distributions of each parameter, while the off-diagonal plots show the joint distributions, indicating a strong correlation between parameters. It is visible that Pantheon+SH0ES data generally provide tighter constraints compared to BAO+CC, with significant overlap in parameter space.}
    \label{fig:rhovsothers}
\end{figure}

The order of difference between pressure and source terms yields a simplified expression for the bounce action:
\begin{equation}
    B_{\text{}} = B_0 + B_2 = - \hbar w_{\text{eff}}(t) - \frac{\hbar}{2}(\partial_0 \phi_f )^2 + \hbar V_f (\phi_f)
\end{equation}Or equivalently:
\begin{equation}
     B_{\text{}} = -\hbar\left(  w_{\text{eff}}(t) + \frac{1}{2}M_{p}^2 \right)
\end{equation}
The effective EoS \((w_{\text{eff}})\) incorporating decay processes using the Breit-Wigner function becomes:
\begin{equation}
w_{\text{eff}}(t) = -\ln \left[ \left(\frac{2 \pi}{\epsilon^\alpha}\right) \frac{\Gamma}{A} \exp(\lambda \phi(t)) \exp(E_0 - E_R) \rho_{de} \right]
\end{equation}where we find that the EoS is directly proportional to $\rho_{de}$ and hence $V_{\text{eff}}(\phi)$. Refer to Fig. \ref{fig:bestfit1} and Fig. \ref{fig:rhovsothers} for a comparison. 

If \( \rho_{de} > 1 \), this corresponds to a high energy density state, which could be linked to an early universe scenario where effective potential dominates. In this state, the universe could be in an excited or unstable false vacuum with a significant potential for energy absorption. This can be mitigated by setting a lower bound on the magnitude of the derivative terms in the potential \( V_{\text{eff}}(\phi) \):
\begin{equation}
    \left(\frac{\partial \Lambda' (\phi)}{\partial \phi}\right)^2 \geq \frac{3}{2}\left(\frac{\partial \phi_f}{\partial a}\right)^{-2}
\end{equation}
ensuring that the system evolves towards a stable configuration.

\begin{figure}[H]
    \centering
    \includegraphics[width=0.7\linewidth]{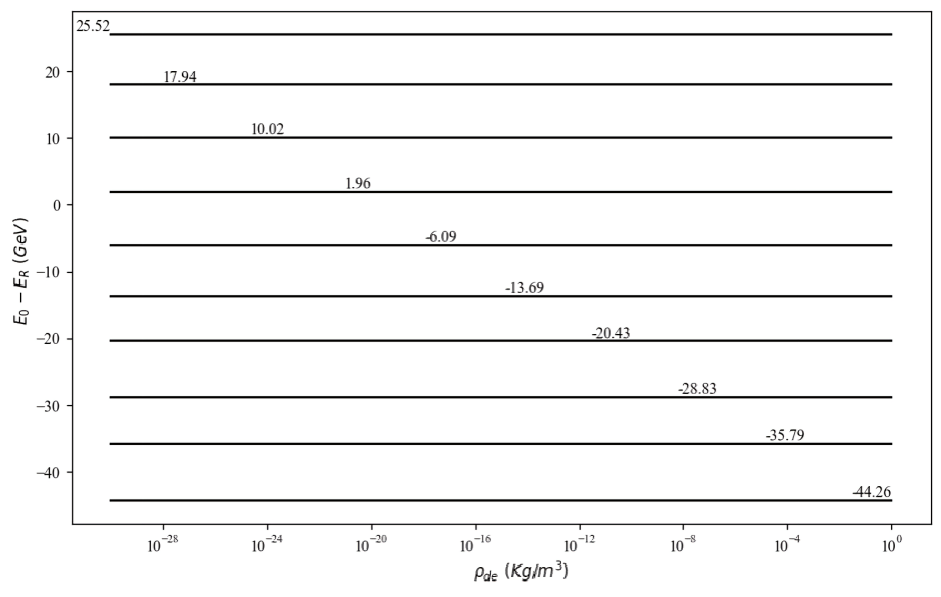}
    \caption{The relationship between \(\rho_{de}\) and \(E_0 - E_R\). As \(\rho_{de}\) decreases, \(E_0 - E_R\) shifts from negative to positive, indicating a transition to a more stable energy configuration.}
    \label{fig:energylevel}
\end{figure}

Conversely, when \( \rho_{de} < 1 \), the energy density is lower, indicative of a later universe where dark energy density has decreased. This condition suggests a transition towards a true vacuum state, where the scalar field decays from an unstable false vacuum to a more stable configuration, releasing energy ( illustrated in Fig. \ref{fig:energylevel}.) that may contribute to the matter density through processes like reheating or particle production \citep{hashiba2021particle}.

The variation in $B_{\text{}}$ with respect to scale factor and scalar field associated with false vacuum becomes:\begin{equation}
a(t)\left(\frac{\partial B_{\text{}}}{{\partial a}}\right) = 0, \quad \phi_f \left(\frac{\partial B_{\text{}}}{{\partial \phi_f}}\right) = 0
\end{equation} For the first bounce term, we start with equation (48):
\begin{equation}
    \frac{\partial B_{\text{}}}{\partial a} = -\hbar \frac{\partial w_{\text{eff}}(t)}{\partial a} - \frac{\hbar}{2} \frac{\partial (\partial_0 \phi_f)^2}{\partial a} + \hbar \frac{\partial V_f (\phi_f)}{\partial a}
\end{equation} Assuming that \(\left(\frac{2 \pi}{\epsilon^\alpha}\right)\), \(\frac{\Gamma}{A}\), \(\exp(\lambda \phi(t))\), and \(\exp(E_0 - E_R)\) are constants with respect to \(a(t)\), we obtain:
\begin{equation}
    a(t) \left[ - \hbar \left( - \frac{1}{\rho_{de}} \frac{\partial \rho_{de}}{\partial a} \right) - \hbar (\partial_0 \phi_f) \frac{\partial (\partial_0 \phi_f)}{\partial a} + \hbar \frac{\partial V_f}{\partial \phi_f} \frac{\partial \phi_f}{\partial a} \right] = 0
\end{equation}Neglecting the kinetic term at late time acceleration simplifies the expression to:
\begin{equation}
      \frac{1}{\rho_{de}} \frac{\partial \rho_{de}}{\partial a}  +  \frac{\partial V_f}{\partial \phi_f} \frac{\partial \phi_f}{\partial a} = 0
\end{equation} which can be rewritten as:
\begin{equation}
    \left(\frac{1}{\rho_{de}} \frac{\partial \rho_{de}}{\partial \phi_f} + \frac{\partial V_f}{\partial \phi_f}\right) \frac{\partial \phi_f}{\partial a} = 0
\end{equation}Since $\frac{\partial \phi_f}{\partial a} \neq 0$, we have \begin{equation}
    \frac{1}{\rho_{de}} \frac{\partial \rho_{de}}{\partial \phi_f} + \frac{\partial V_f}{\partial \phi_f} =0
\end{equation} This equation describes the normalized rate of change of $\rho_{de}$ with respect to $\phi_f$. In this case, any change in dark energy density with respect to a $\phi_f$ is counterbalanced by the corresponding change in the (false vacuum) potential $V_f$. While not a strict equivalence, an exponential change in the scalar field could impact $\rho_{de}$, which expands asymptotically with the slow decay of false vacuum potential.

Similarly, for the second bounce term, starting from equation (49): 
\begin{equation}
    \frac{\partial B_{\text{}}}{\partial \phi_f} = - \hbar \frac{\partial w_{\text{eff}}(t)}{\partial \phi_f} 
\end{equation} From section (6), we know that \(\left(\frac{2 \pi}{\epsilon^\alpha}\right)\), \(\frac{\Gamma}{A}\), and \(\exp(E_0 - E_R)\) are field independent parameters of $V_{\text{eff}}$. Thus, we can write:
\begin{equation}
    - \phi_{f} \left[  \left( \lambda \frac{\partial \phi(t)}{\partial \phi_f} + \frac{1}{\rho_{de}} \frac{\partial \rho_{de}}{\partial \phi_f} \right) \right] = 0
\end{equation} Since $\phi_f \neq 0$, we get:
\begin{equation}
    \lambda \frac{\partial \phi(t)}{\partial \phi_f} + \frac{1}{\rho_{de}} \frac{\partial \rho_{de}}{\partial \phi_f}  =0
\end{equation} which shows a dynamic equilibrium condition where variations in \(\phi(t)\) are balanced by variations in \(\rho_{de}\) when $\lambda < 0$. 

\begin{figure}[H]
    \centering
    \subfloat[ With constraints that set the lower bound at \(\lambda < 0.1\) and \(0.3 < \alpha < 0.4\), $V_{\text{eff}}$ slow roll in the range \(-0.8 < w_{\text{eff}} < -0.4\) shown in Fig. \ref{fig:volume}]{\includegraphics[width=0.7\textwidth]{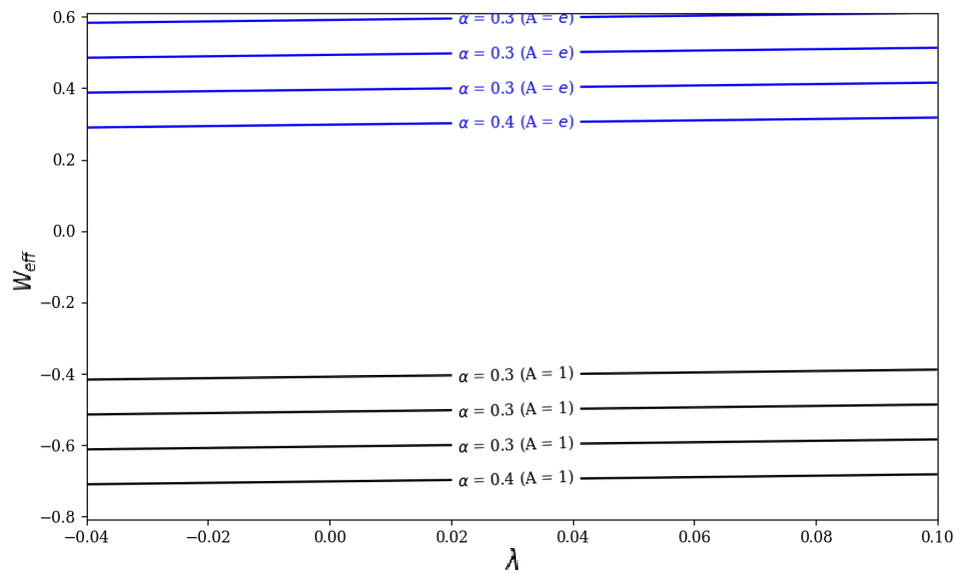} \label{fig:weff}}
    \hfill
    \subfloat[The effective potential \(V_{\text{eff}}\) changes during the transition. The potential reaches its maximum when \(\mathcal{V}_{\text{Bub}} \sim 1\) is at its minimum value, stabilizing the false vacuum.]{\includegraphics[width=0.7\textwidth]{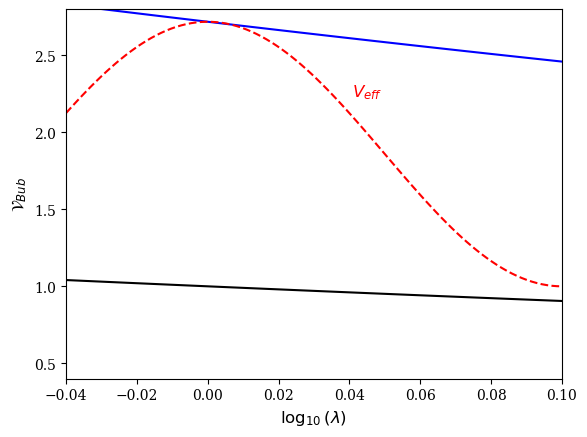} \label{fig:volume}}
    \caption{}
    \label{fig:combined_figures_2}
\end{figure}

The conservation equation of a system involving both true and false vacuum states in a cosmological context becomes:
\begin{align}
    w_{\text{eff}}^{'} &+ (\partial_0 \phi_f) \frac{\partial}{\partial a}(\partial_0 \phi_f) - V_{f}^{'}(\phi_f) - \frac{\partial}{\partial a} \ln \left(\frac{2 \pi}{\epsilon^\alpha}\right) \frac{\Gamma}{A} \exp(E_0 -E_R) \nonumber \\
    &+ \frac{\partial}{\partial a} \ln (\rho_{de}) - \frac{\partial}{\partial a} \ln (-V_t (\phi_t (t))) = 0
\end{align} where $'$ symbol indicates the derivative with respect to scale factor a. To approximate the interaction parameter with respect to Euclidean time, we denote the first and second-order derivatives as:
\begin{equation}
  \phi_f^{'} = \frac{d \phi_f}{a d\tau} = \frac{\phi_f^{'}}{a}, \quad \text{and} \quad \phi_{f}^{''} = \frac{a \phi_{f}^{''} - \phi_{f}^{'}}{a^2}  
\end{equation}
We obtain the order of parameters as:
\begin{equation}
  \ln \left( \left(\frac{2 \pi}{\epsilon^\alpha}\right) \frac{\Gamma}{\mathcal{V}} \exp(E_0 - E_R)\right) \sim c_1, \quad \text{and} \quad \ln (-V_t \phi_t (t)) \sim c_2  
\end{equation}where we can consider these bounds (refer to Fig. \ref{fig:combined_figures_2}) based on the relative magnitudes of parameters \(c_1\) and \(c_2\), which are derived from different contributions to the effective potential and dark energy density.  

When \( c_1 > c_2 \), the energy transition is dominated by the effective potential, leading to a rapid growth in effective potential. In contrast, if \( c_1 < c_2 \),  the effective potential grows more slowly, leading to a gradual transition in the dark energy density. The volume of the bubble does not grow exponentially; rather, the system remains stable, representing a stable class of quintessence models, which we will discuss later in the section.

When $c_1 = c_2$, the background repulsion matches the inward gravitational attraction from the true vacuum. This case is similar to the one studied by \citep{sivaram2011dieterici} when the values of $c_1$ and $c_2$ are on the order of $\sim 10^{31} \text{GeV} $ (less than the required from equation (36)). Then:
\begin{equation}
    w_{\text{eff}}^{'} - V_{f}^{'}(\phi_f) + \frac{\partial}{\partial a} \ln (\rho_{de}) = 0
\end{equation} We substitute the expression for $ w_{\text{eff}}^{'}$ to get :
\begin{equation}
   - \frac{1}{\rho_{de}} \frac{\partial \rho_{de}}{\partial a} - V_{f}^{'}(\phi_f) + \frac{\partial}{\partial a} \ln (\rho_{de}) = 0
\end{equation} which gives:
\begin{equation}
    - V_{f}^{'}(\phi_f) =0
\end{equation} Or phenomenologically equivalent to:
\begin{equation}
      V_{\text{eff}}^{'}(\phi_f) = -Q
\end{equation}We have included an additional term $Q$ for interaction, which can be interpreted as the energy transfer rate between dark energy and the false vacuum. Else equation (66) corresponds to a local extremum of the potential function. Moreover, it is also visible that the behaviour of \(Q\) directly impacts key cosmological observables. Depending on the stability and form of \(V_{\text{eff}}\), the late-time evolution of the Hubble parameter may exhibit distinct behaviours, as shown in Table \ref{table:Q_effects}.

\begin{table}[h!]
\centering
\begin{tabular}{|>{\raggedright\arraybackslash}p{1.8cm}|>{\raggedright\arraybackslash}p{3cm}|>{\raggedright\arraybackslash}p{3.2cm}|>{\raggedright\arraybackslash}p{3.5cm}|}
\hline
\hspace{5mm}\textbf{\( Q \)} & \hspace{9mm}\textbf{\( \rho_{de} \)} & \hspace{9mm}\textbf{\( w_{\text{eff}} \)} & \hspace{13mm}\textbf{\( H(t) \)} \\ 
\hline
Small and Smooth \( Q \) & Gradual energy transfer stabilizes \( \rho_{de} \), ensuring smooth field evolution. &  \( w_{\text{eff}} \to -1 \): Mimics a cosmological constant, sustaining acceleration. & Smooth decrease in \( H(t) \), consistent with late-time acceleration and observational constraints. \\ 
\hline
Rapidly Varying \( Q \) &  Oscillations or instabilities in \( \rho_{de} \), disrupting stability and smooth evolution. &  \( w_{\text{eff}} \) deviates from \(-1\): Causes oscillatory or non-monotonic behavior. & Fluctuations in \( H(t) \), potentially inconsistent with late-time acceleration. \\ 
\hline
Large \( Q \) & Rapid depletion of \( \rho_{de} \); dark energy decays too quickly. &  \( w_{\text{eff}} > -1 \): Field evolves prematurely, reducing dark energy dominance. &  \( H(t) \) slows early, leading to insufficient late-time acceleration. \\ 
\hline
Small and Rapidly Oscillating \( Q \) & Alternating energy transfer causes oscillatory \( \rho_{de} \), with temporary instability. & Periodic fluctuations in \( w_{\text{eff}} \); may average close to \(-1\) over time. & Oscillatory \( H(t) \); may stabilize over long timescales. \\ 
\hline
\( Q \approx 0 \) &  No energy transfer; field remains trapped in the false vacuum. & \( w_{\text{eff}} \) remains constant or undefined; acceleration ceases. & \( H(t) \) stagnates, failing to produce late-time acceleration. \\ 
\hline
\end{tabular}
\caption{\small The table above qualitatively summarizes how the different behaviours of \( Q \) influence late-time cosmic evolution. To avoid eternally stable de Sitter vacua, \( Q \) values should comply with Swampland constraints while enabling late-time cosmic acceleration and a transition consistent with \(\Lambda\)-CDM predictions. } 
\label{table:Q_effects}
\end{table}
Equation (67) can be rewritten in standard form:
\begin{equation}
    \dot{\rho_{\text{de}}} = -Q, \quad  \dot{\rho_m} + 3 H \rho_m = Q, \quad  \dot{\rho_r} + 4 H \rho_r = 0
\end{equation} where $\dot{\rho_m}$, $\dot{\rho_r}$ and $\dot{\rho_{de}}$ denotes the time derivative of matter, radiation, and dark energy density, respectively. To be more precise, for the interaction term:
\begin{equation}
    \frac{1}{\rho_{de}} \frac{\partial \rho_{de}}{\partial a}  = -\frac{Q}{\rho_{de}},
\end{equation}
where \( Q \) is the interaction term defined by:
\begin{equation}
    Q = -\Gamma \rho_{de}
\end{equation}
\(\Gamma\) represents the decay width or decay rate of the false vacuum, and \( \rho_{de} \) the dark energy density. Thus, the equation for the evolution of the potential becomes:
\begin{equation}
    \frac{\partial V_{\text{eff}}}{\partial \phi_f} \frac{\partial \phi_f}{\partial a} = -\Gamma \rho_{de}
\end{equation} Note that in the absence of interaction rate term  \( Q \), no forces will act in the potential to drive the field. This means that $\rho_{de}$ is independent of the dynamics of the false vacuum potential at a local minimum or maximum. Hence, with the addition of  \( Q \),  the force due to the potential gradient is exactly counterbalanced by the interaction term. This makes the system stable when these forces are in equilibrium.

\begin{figure}[H]
    \centering
    \includegraphics[width = 0.8\linewidth]{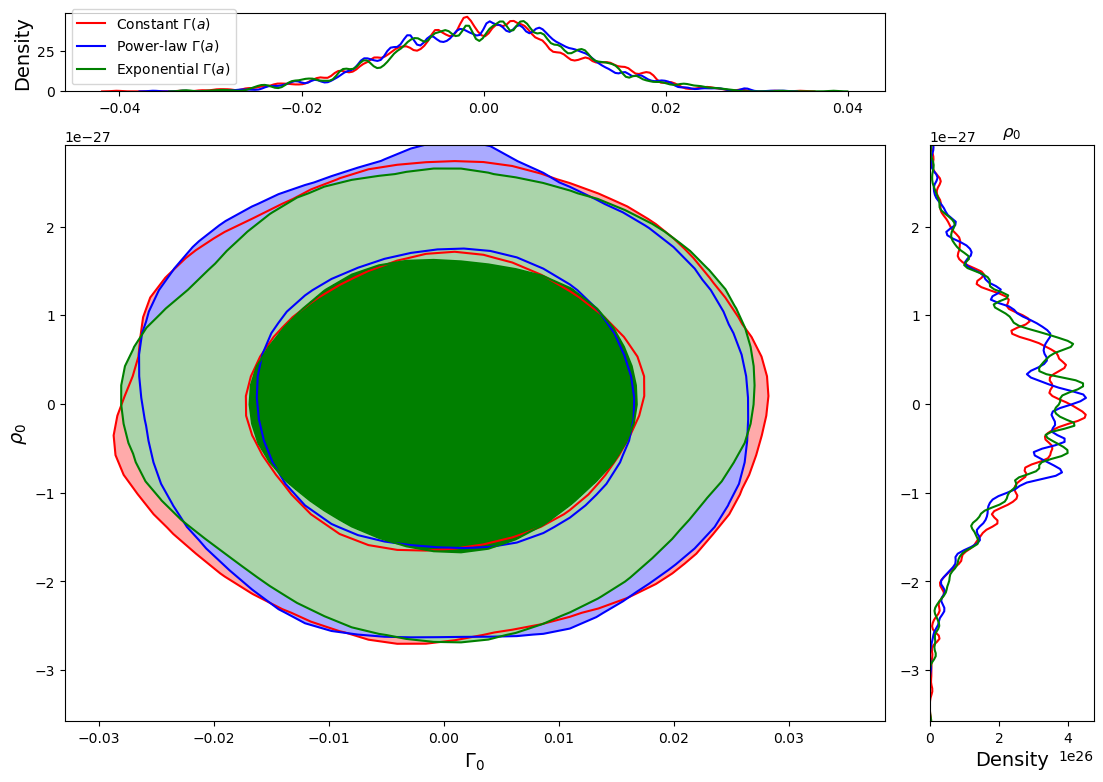}
    \caption{\small Compares the joint and marginal posterior distributions of $\Gamma_0$ and $\rho_0$ for three cosmological models with contours representing 68\% and 95\% confidence levels using combined Panthen+SHOES and BAO+CC data. The marginal distributions of \( \Gamma_0 \) (top) and \( \rho_0 \) (right) are visualized using Kernel Density Estimation (KDE) for arbitrary values of other model parameters. The top plot shows the distribution of \( \Gamma_0 \), while the plot on the right shows the \( \rho_0 \) distribution.  }
    \label{fig:threemodels}
\end{figure}

The general solution for equation (71) with arbitrary functions \( \Gamma(a) \) and \( \rho_{de}(a) \) is:
\begin{equation}
    V_{\text{eff}}(a) = -\int \Gamma(a) \rho_{de}(a) \ da + C
\end{equation} with C being an integration constant. Depending on the specific forms of $\Gamma (a)$ and $\rho_{\textit{de}}(a)$, we can derive more explicit solutions: 
\begin{enumerate}
    \item  If both \( \Gamma(a) = \Gamma_0 \) and \( \rho_{de}(a) = \rho_0 \) are constants, the effective potential \( V_{\text{eff}}(a) \) varies linearly with \( a \). In this scenario:
\begin{equation}
    V_{\text{eff}}(a) = \Gamma_0 \rho_0 a + C
\end{equation}The model implies a steady force in the expansion, similar to the dynamics of the $\Lambda$-CDM model where dark energy dominates and drives the acceleration of the universe. 
\item If \( \Gamma(a) \) and \( \rho_{de}(a) \) follow a power-law form, with \( \Gamma(a) = \Gamma_0 a^n \) and \( \rho_{de}(a) = \rho_0 a^m \), the potential scales as a power of \( a \):
\begin{equation}
\begin{split}
V_{\text{eff}}(a) &= \Gamma_0 \rho_0 \int a^{n+m} \, da + C \\
                  &= \frac{\Gamma_0 \rho_0}{n+m+1} a^{n+m+1} + C
\end{split}
\end{equation}where \( n + m + 1 >0 \). The power-law dependence indicates a rapid change in the dark energy density, especially if \( n + m \) is large, akin to the behaviour during inflation.
\item For an exponential dependence, where \( \Gamma(a) = \Gamma_0 \exp(\alpha a) \) and \( \rho_{de}(a) = \rho_0 \exp(\beta a) \), the potential increases exponentially with \( a \):
\begin{equation}
\begin{split}
V_{\text{eff}}(a) &= \Gamma_0 \rho_0 \int \exp((\alpha + \beta) a) \, da + C \\
                  &= \frac{\Gamma_0 \rho_0}{\alpha + \beta} \exp((\alpha + \beta) a) + C
\end{split}
\end{equation}where \( \alpha + \beta \neq 0 \). Such potentials are common in scalar dark energy models, specifically those inspired by quintessence scalars. Compared to the above two, this model suggests a positive contribution to the effective potential with \(\Gamma_0\) being positive and better constrained with observed data, shown in Fig \ref{fig:threemodels}.
\end{enumerate}

\section{Future Prospects}
The quintessence field and its associated false vacuum state share similarities with old inflation models, where inflation ends in a first-order phase transition, in contrast to the second-order slow-roll phase transition of new inflation \citep{PhysRevLett.48.1220}. In these scenarios, the scalar field is initially trapped in a false vacuum state (a local minimum of the potential), which eventually decays to a lower-energy true vacuum state. The implications of this mechanism are profound, as it not only provides a plausible end to the inflationary epoch but also opens up the possibility of explaining certain features of the current accelerated expansion through similar dynamical processes. However, associating a dark energy model with an inflationary model has to be done thoughtfully, as it can lead to inhomogeneities and potential issues with reheating that are inconsistent with current observations.

While our analysis has primarily focused on semi-classical conditions, the role of quantum corrections cannot be overlooked. A more detailed theoretical treatment of the quintessence field could provide deeper insights into the stability and evolution of scalar potential, particularly in the context of vacuum fluctuations and tunnelling processes. This is particularly relevant in light of our findings, which suggest that the current energy configuration of our universe might be a result of a slow-evolving scalar rather than being arbitrary. The proximity of  \( \rho_{de} \) to its current value implies that  $E_0>E_R$ at present, necessitating an additional initial term in the equation of motion for the scalar to prevent late-time de Sitter vacua. Interestingly, the early dark energy hypothesis - where a new form of dark energy \citep{niedermann2024new, niedermann2021new, herold2023resolving, hill2020early, kamionkowski2023hubble}, active at early times and dilutes away around the matter-radiation equality - offers a potential solution to avoid late-time cosmological catastrophes.

However, the stability of the quintessence scalar in the early times remains an open question. Especially considering the impact of effective Higgs self-coupling, which requires a field theory framework. As the Higgs field interacts with Standard model particles, there is a possibility that thermal fluctuations might trigger the decay of false vacuum \citep{Sher1989, Arnold1991, Espinosa1995}. Addressing this issue may require exploring the link between the equation of state (EoS) parameter and effective field theory parameters within a multi-axion framework\citep{niedermann2024new}. Higher-order terms in the Higgs potential could also induce electroweak symmetry breaking during the early universe via the freeze-in mechanism   \citep{PhysRevLett.116.101302}. These factors present significant challenges to the model as currently formulated, as it does not fully account for these effects.

The model also requires careful fine-tuning of parameters, particularly in ensuring the stability of the false vacuum and preventing premature decay. Future work should focus on exploring the parameter space more thoroughly and considering possible extensions to the model that could naturally alleviate some of these fine-tuning requirements. A natural extension would be exploring the phenomenological interaction between the scalar field (often associated with dark energy) and dark matter. Since it is still unknown to what extent the dark sector remains consistent with the concordance model, it is compelling to think that similar transitions can occur in the dark sector as studied in \citep{di2017can, di2020interacting, di2020nonminimal, baer2015dark, wang2016probing, wang2016dark, muka2017Falsevaccum}. If the quintessence field undergoes a phase transition, it might be accompanied by a corresponding transition in the dark matter sector \citep{lepe2016interacting, banihashemi2020phase, brandenberger2019unified}. This could lead to observable consequences such as the generation of dark matter substructures or changes in dark matter annihilation rates, which could be probed by indirect detection experiments \citep{perez2020status, klasen2015indirect, slatyer2018indirect}.

\section{Conclusion}
In this paper, we studied a quintessence field initially trapped in a metastable false vacuum state, analyzing its evolution and stability at late times while considering the Swampland criteria. Swampland criteria are appropriate for the future possibility of modelling interactions, assuming that gravity stabilizes false vacuum decay. We established new bounds on the energy-momentum tensor, considering a fixed pressure term for the false vacuum transition. This setup allowed the scalar field to evolve within flat spacetime under fixed boundary conditions, effectively mimicking the behaviour of a running cosmological constant through the gradual decay of the false vacuum. Our findings suggest that, contrary to gravity-stabilized scenarios, certain semi-classical conditions can stabilize false vacuum decay, even within the Swampland constraints.

The Swampland criteria impose stringent requirements on the scalar field potential, demanding steep potentials and limited field excursions. We demonstrated that a scalar field potential with a local minimum at a positive value could satisfy these criteria, preventing the formation of an eternally stable de Sitter vacuum. This result is significant as it aligns with the behaviour of a slow-rolling effective potential \(V_{\text{eff}}\) within the range \(0.1 > \lambda > -0.04\), spanning from \(A \sim 1\) to \(A = 2.718\), with maximum stabilization occurring when \(\mathcal{V}_{\text{Bub}} \sim 1.31\).

The evolution of the quintessence field, as described by differential equation (26) and potential such as \(V(\phi_c) = \frac{1}{2} m^2 \phi^2\), suggests a slow-roll process where the damping term \(3H\dot{\phi}\) dominates, leading to a slowly decreasing \(\phi(t)\). This mirrors the exponential form commonly seen in inflationary models, where the potential term \(V(\phi_c(t))\) primarily drives the expansion. For a quintessence field, the potential \(V_t(\phi_c(t))\) might take a power-law form, typical in some scalar dark energy models such as K-essence and tachyonic dark energy. While exponential forms are often found in string theory compactifications, it is worth mentioning that the primary perturbative corrections that occur after supersymmetry breaking tend to elevate flat directions in the moduli potential, resulting in steep runaway behaviour.

The effective EoS parameter, derived from the source term \(\Lambda' (\phi) = E_0 - E_R\), invokes the realization that the current energy configuration is not arbitrary. A large positive value of \((E_0 - E_R)\) indicates a slow evolution of the quintessence field from a false vacuum state, while a negative value suggests an increase in dark energy density at the present epoch. Suppose that dark energy evolved from a higher value than its present value. This supports the notion that the current vacuum state is finely tuned, where it no longer depends on the initial meta-state, the quintessence has primarily evolved.

The effective EoS was further tested against observational data, using separate and combined analysis using Pantheon+ SHOES and BAO+CC data.  In the combined analysis (Fig. \ref{fig:bestfit1}), we obtained maximum likelihood parameters of \(A \approx 2.41\), \(\alpha \approx 0.32\), \(w_{\text{eff}} \approx -0.7\), \(\lambda \approx -0.03\), and \((E_0 - E_R) \approx 19\) GeV, fits well within a physically plausible range. The contour plots generally show weak to moderate correlations, indicating that while some parameters exhibit dependencies (e.g., \(E_0 - E_R\) with \(\lambda\) and \(w_{\text{eff}}\)), many parameters independently contribute to the model without heavily influencing others. Whereas the separate analysis (Fig. \ref{fig:rhovsothers}) shows slightly different values for model parameters when tested for $\rho_{de}$ with the above-mentioned model parameters. Most parameters show peaks that suggest minimal deviation from the standard $\Lambda$-CDM model, indicating consistency between the two data sets.  

Moreover, the stability of the quintessence field was examined under the (phenomenological) condition \(-V_{f}^{'}(\phi_f) = Q\) and \(-V_{f}^{'}(\phi_f) = 0\). When \(-V_{f}^{'}(\phi_f) = Q\), the potential gradient is counterbalanced by an interaction term \(Q\), indicating that the field dynamics are influenced by both the potential and interaction terms, leading to new bounds on the quintessence field. On the other hand, when \(-V_{f}^{'}(\phi_f) = 0\), the gradient of the false vacuum potential is zero, characterizing a stable configuration where the potential does not change with \(\phi_f\). This typically signifies an equilibrium state or a minimum, where the field's dynamics are governed by the slow decay of the false vacuum rather than by a changing potential.

However, it is important to note several limitations of our model. The assumption of the false vacuum in a stable state may not hold for phantom energies, potentially leading to inconsistencies with model parameters. The model's dependence is oversimplified based on a single field parameter \(\lambda\), which may also present fine-tuning challenges and require more computational time. A non-zero \(\lambda\) suggests that the dynamics of the quintessence field are directly influenced by the state of the false vacuum, which may require revision considering additional interactions at small scales. Furthermore, the dynamics of the false vacuum transition could be affected by multiple quintessence fields or by coupling to other fields, aspects not addressed in this study. The exclusion of other dark matter particles and the lack of consideration for quantum gravity effects also limit the model's applicability.

While our study offers a novel perspective on the evolution of a simple scalar dark energy model with exponential potential, several challenges remain. Future research should explore numerical simulations and advanced parameter estimation techniques to investigate the dynamics and observational implications of non-exponential potentials. By integrating observational constraints from next-generation surveys with stability and theoretical analyses, we can validate or potentially rule out these non-exponential potentials as viable alternatives to the exponential form. Although these possibilities are beyond the current scope of our study, they offer promising directions for future investigation. These steps will help refine the model and provide a more comprehensive understanding of the dark sector and its role in the evolution of the universe.

\section{Declaration of competing interest}
\begin{enumerate}
    \item Author Contributions: All authors contributed equally to the preparation of the manuscript.
    \item Funding: Seed money scheme, Sanction No. CU-ORS-SM-24/29.
    \item Data Availability Statement: Not applicable.
    \item Conflicts of Interest: The authors declare no conflict of interest.
\end{enumerate}
\section{Acknowledgments}
The authors would like to express their gratitude to Christ (Deemed to be University) for funding this research under the Seed Money Project.

\bibliographystyle{unsrtnat}  
\bibliography{ref}            

\begin{thebibliography}{105}
\providecommand{\natexlab}[1]{#1}
\providecommand{\url}[1]{\texttt{#1}}
\expandafter\ifx\csname urlstyle\endcsname\relax
  \providecommand{\doi}[1]{doi: #1}\else
  \providecommand{\doi}{doi: \begingroup \urlstyle{rm}\Url}\fi

\bibitem[Perlmutter and et~al.(1999)]{perlmutter1999measurements}
S.~Perlmutter and et~al.
\newblock Measurements of $\omega$ and $\lambda$ from 42 high redshift supernovae.
\newblock \emph{Astrophysical Journal}, 517:\penalty0 565--586, 1999.

\bibitem[Riess and et~al.(1998)]{riess1998observational}
Adam~G. Riess and et~al.
\newblock Observational evidence from supernovae for an accelerating universe and a cosmological constant.
\newblock \emph{The Astronomical Journal}, 116\penalty0 (3):\penalty0 1009, 1998.

\bibitem[Popolo and Delliou(2016)]{delpopolo2016}
A.~Del Popolo and M.~Le Delliou.
\newblock Small scale problems of the $\lambda$cdm model: A short review.
\newblock \emph{arXiv preprint arXiv:1606.07790}, 2016.

\bibitem[Turner(2018)]{turner2018}
M.~S. Turner.
\newblock $\lambda$cdm: Much more than we expected, but now less than what we want.
\newblock \emph{Found Phys}, 48:\penalty0 1261–1278, 2018.
\newblock \doi{10.1007/s10701-018-0178-8}.

\bibitem[Weinberg(2001)]{weinberg2001}
S.~Weinberg.
\newblock The cosmological constant problems.
\newblock In D.~B. Cline, editor, \emph{Sources and Detection of Dark Matter and Dark Energy in the Universe}. Springer, Berlin, Heidelberg, 2001.
\newblock \doi{10.1007/978-3-662-04587-9_2}.

\bibitem[Lombriser(2019)]{lombriser2019cosmological}
Lucas Lombriser.
\newblock On the cosmological constant problem.
\newblock \emph{Physics Letters B}, 797:\penalty0 134804, 2019.

\bibitem[Padilla(2015)]{padilla2015lectures}
Antonio Padilla.
\newblock Lectures on the cosmological constant problem.
\newblock \emph{arXiv preprint arXiv:1502.05296}, 2015.

\bibitem[Ng(1992)]{ng1992cosmological}
Y~Jack Ng.
\newblock The cosmological constant problem.
\newblock \emph{International Journal of Modern Physics D}, 1\penalty0 (01):\penalty0 145--160, 1992.

\bibitem[Barrow and Shaw(2011)]{barrow2011new}
John~D Barrow and Douglas~J Shaw.
\newblock New solution of the cosmological constant problems.
\newblock \emph{Physical Review Letters}, 106\penalty0 (10):\penalty0 101302, 2011.

\bibitem[Garriga and Vilenkin(2001)]{garriga2001solutions}
J~Garriga and Alexander Vilenkin.
\newblock Solutions to the cosmological constant problems.
\newblock \emph{Physical Review D}, 64\penalty0 (2):\penalty0 023517, 2001.

\bibitem[Carroll(2001)]{carroll2001cosmological}
Sean~M Carroll.
\newblock The cosmological constant.
\newblock \emph{Living reviews in relativity}, 4\penalty0 (1):\penalty0 1--56, 2001.

\bibitem[Padmanabhan(2003)]{padmanabhan2003cosmological}
Thanu Padmanabhan.
\newblock Cosmological constant—the weight of the vacuum.
\newblock \emph{Physics reports}, 380\penalty0 (5-6):\penalty0 235--320, 2003.

\bibitem[Sahni(2002)]{sahni2002cosmological}
Varun Sahni.
\newblock The cosmological constant problem and quintessence.
\newblock \emph{Classical and Quantum Gravity}, 19\penalty0 (13):\penalty0 3435, 2002.

\bibitem[Barr and Seckel(2001)]{barr2001cosmological}
Stephen~M Barr and D~Seckel.
\newblock Cosmological constant, false vacua, and axions.
\newblock \emph{Physical Review D}, 64\penalty0 (12):\penalty0 123513, 2001.

\bibitem[Bousso et~al.(2015)Bousso, Harlow, and Senatore]{bousso2015inflation}
Raphael Bousso, Daniel Harlow, and Leonardo Senatore.
\newblock Inflation after false vacuum decay: observational prospects after planck.
\newblock \emph{Physical Review D}, 91\penalty0 (8):\penalty0 083527, 2015.

\bibitem[Blau et~al.(1987)Blau, Guendelman, and Guth]{blau1987dynamics}
Steven~K Blau, Eduardo~I Guendelman, and Alan~H Guth.
\newblock Dynamics of false-vacuum bubbles.
\newblock \emph{Physical Review D}, 35\penalty0 (6):\penalty0 1747, 1987.

\bibitem[Krauss and Dent(2008)]{krauss2008late}
Lawrence~M Krauss and James Dent.
\newblock Late time behavior of false vacuum decay: Possible implications for cosmology<? format?> and metastable inflating states.
\newblock \emph{Physical Review Letters}, 100\penalty0 (17):\penalty0 171301, 2008.

\bibitem[Rafelski and Birrell(2015)]{rafelski2015dynamical}
Johann Rafelski and Jeremiah Birrell.
\newblock Dynamical emergence of the universe into the false vacuum.
\newblock \emph{Journal of Cosmology and Astroparticle Physics}, 2015\penalty0 (11):\penalty0 035, 2015.

\bibitem[Fischler et~al.(1990)Fischler, Morgan, and Polchinski]{fischler1990quantization}
Willy Fischler, Daniel Morgan, and Joseph Polchinski.
\newblock Quantization of false-vacuum bubbles: A hamiltonian treatment of gravitational tunneling.
\newblock \emph{Physical Review D}, 42\penalty0 (12):\penalty0 4042, 1990.

\bibitem[Bronnikov(2001)]{bronnikov2001spherically}
Kirill~A Bronnikov.
\newblock Spherically symmetric false vacuum: no-go theorems and global structure.
\newblock \emph{Physical Review D}, 64\penalty0 (6):\penalty0 064013, 2001.

\bibitem[Kost et~al.(2022)Kost, Shin, and Terada]{kost2022massless}
Jeff Kost, Chang~Sub Shin, and Takahiro Terada.
\newblock Massless preheating and electroweak vacuum metastability.
\newblock \emph{Physical Review D}, 105\penalty0 (4):\penalty0 043508, 2022.

\bibitem[Smeenk(2005)]{smeenk2005false}
Chris Smeenk.
\newblock False vacuum: Early universe cosmology and the development of inflation.
\newblock In \emph{The universe of general relativity}, pages 223--257. Springer, 2005.

\bibitem[Sato(1981)]{sato1981first}
Katsuhiko Sato.
\newblock First-order phase transition of a vacuum and the expansion of the universe.
\newblock \emph{Monthly Notices of the Royal Astronomical Society}, 195\penalty0 (3):\penalty0 467--479, 1981.

\bibitem[Mersini-Houghton(2006)]{mersini2006}
Laura Mersini-Houghton.
\newblock The arrow of time forbids a positive cosmological constant $Λ$.
\newblock \emph{arXiv: General Relativity and Quantum Cosmology}, 2006.

\bibitem[Eichhorn et~al.(2024)Eichhorn, Hebecker, Pawlowski, and Walcher]{eichhorn2024absolute}
Astrid Eichhorn, Arthur Hebecker, Jan~M Pawlowski, and Johannes Walcher.
\newblock The absolute swampland.
\newblock \emph{arXiv preprint arXiv:2405.20386}, 2024.

\bibitem[Palti(2019)]{palti2019swampland}
Eran Palti.
\newblock The swampland: introduction and review.
\newblock \emph{Fortschritte der Physik}, 67\penalty0 (6):\penalty0 1900037, 2019.

\bibitem[Obied et~al.(2018)Obied, Ooguri, Spodyneiko, and Vafa]{obied2018}
G.~Obied, H.~Ooguri, L.~Spodyneiko, and C.~Vafa.
\newblock de sitter space and the swampland.
\newblock \emph{arXiv preprint arXiv:1806.08362}, 2018.

\bibitem[Cosemans and Smet(2005)]{cosemans2005stable}
Bert Cosemans and Geert Smet.
\newblock Stable de sitter vacua in n= 2, d= 5 supergravity.
\newblock \emph{Classical and Quantum Gravity}, 22\penalty0 (12):\penalty0 2359, 2005.

\bibitem[Espinosa(2020)]{espinosa2020stabilizing}
JR~Espinosa.
\newblock The stabilizing effect of gravity made simple.
\newblock \emph{Journal of Cosmology and Astroparticle Physics}, 2020\penalty0 (07):\penalty0 061, 2020.

\bibitem[Cveti{\v{c}} et~al.(1993)Cveti{\v{c}}, Griffies, and Rey]{cvetivc1993non}
Mirjam Cveti{\v{c}}, Stephen Griffies, and Soo-Jong Rey.
\newblock Non-perturbative stability of supergravity and superstring vacua.
\newblock \emph{Nuclear Physics B}, 389\penalty0 (1):\penalty0 3--24, 1993.

\bibitem[Gliner(1965)]{gliner1965}
E.~B. Gliner.
\newblock Algebraic properties of the energy-momentum tensor and vacuum-like states of matter.
\newblock \emph{ZhTF}, 49:\penalty0 542–548, 1965.
\newblock In Russian. English transl.: Sov. Phys. JETP 1966, 22, 378.

\bibitem[Gliner(1970)]{gliner1970}
E.~B. Gliner.
\newblock Vacuum-like state of medium and friedmann’s cosmology.
\newblock \emph{Akademiia Nauk SSSR Doklady}, 192:\penalty0 771–774, 1970.
\newblock In Russian. Engl. transl: Sov. Phys. Dokl. 1970, 15, 559.

\bibitem[Gliner and Dymnikova(1975)]{gliner1975}
E.~B. Gliner and I.~G. Dymnikova.
\newblock A nonsingular friedmann cosmology.
\newblock \emph{Pisma Astron. Zhurnal}, 1:\penalty0 7, 1975.
\newblock In Russian. Engl. transl.: Sov. Astron. Lett. 1975, 1, 93.

\bibitem[Starobinsky(1979)]{starobinsky1979}
A.~A. Starobinsky.
\newblock Spectrum of gravitational background radiation and initial state of the universe.
\newblock \emph{ZhTF Pisma}, 30:\penalty0 719–723, 1979.
\newblock In Russian. Engl. Transl.: JETP Lett. 1979, 30, 682.

\bibitem[Guth(1981)]{guth1981}
A.~H. Guth.
\newblock Inflationary universe: A possible solution to the horizon flatness problems.
\newblock \emph{Physical Review D}, 23:\penalty0 347--356, 1981.

\bibitem[Wald(1983)]{wald1983}
R.~M. Wald.
\newblock Asymptotic behavior of homogeneous cosmological models in the presence of a positive cosmological constant.
\newblock \emph{Phys. Rev. D}, 28:\penalty0 2118--2120, 1983.

\bibitem[Juárez-Aubry(2019)]{juarez-aubry2019}
B.~A. Juárez-Aubry.
\newblock Semi-classical gravity in de sitter spacetime and the cosmological constant.
\newblock \emph{Phys. Lett. B}, 797:\penalty0 134912, 2019.

\bibitem[Coleman(1977)]{coleman1977}
S.~R. Coleman.
\newblock Fate of the false vacuum: Semiclassical theory.
\newblock \emph{Phys. Rev. D}, 15:\penalty0 2929, 1977.
\newblock \doi{10.1103/PhysRevD.15.2929}.

\bibitem[Coleman and Luccia(1980)]{coleman1980}
S.~R. Coleman and F.~De Luccia.
\newblock Gravitational effects on and of vacuum decay.
\newblock \emph{Phys. Rev. D}, 21:\penalty0 3305, 1980.
\newblock \doi{10.1103/PhysRevD.21.3305}.

\bibitem[Jr. and Coleman(1977)]{callan1977}
C.~G.~Callan Jr. and S.~R. Coleman.
\newblock Fate of the false vacuum. ii. first quantum corrections.
\newblock \emph{Phys. Rev. D}, 16:\penalty0 1762, 1977.
\newblock \doi{10.1103/PhysRevD.16.1762}.

\bibitem[Wetterich(1988)]{wetterich1988cosmology}
C.~Wetterich.
\newblock Cosmology with variable cosmological constant.
\newblock \emph{Nuclear Physics B}, 302\penalty0 (4):\penalty0 668--696, 1988.

\bibitem[Wetterich(2004)]{wetterich2004phenomenological}
C.~Wetterich.
\newblock Phenomenological parameterization of quintessence.
\newblock \emph{Physics Letters B}, 594\penalty0 (1-2):\penalty0 17--22, 2004.
\newblock \doi{10.1016/j.physletb.2004.04.080}.

\bibitem[Stachowski et~al.(2016)Stachowski, Szydlowski, and Urbanowski]{stachowski2016}
Aleksander Stachowski, Marek Szydlowski, and Krzysztof Urbanowski.
\newblock Cosmological implications of the transition from the false vacuum to the true vacuum state.
\newblock \emph{European Physical Journal C}, 77, 2016.
\newblock \doi{10.1140/epjc/s10052-017-4934-2}.

\bibitem[Calmet and El-Menoufi(2017)]{calmet2017}
X.~Calmet and B.~K. El-Menoufi.
\newblock Quantum corrections to schwarzschild black hole.
\newblock \emph{Eur. Phys. J. C}, 77\penalty0 (4):\penalty0 243, 2017.
\newblock \doi{10.1140/epjc/s10052-017-4802-0}.

\bibitem[Masina and Notari(2012{\natexlab{a}})]{masina2012}
I.~Masina and A.~Notari.
\newblock The higgs mass range from standard model false vacuum inflation in scalar-tensor gravity.
\newblock \emph{Phys. Rev. D}, 85:\penalty0 123506, 2012{\natexlab{a}}.

\bibitem[Masina and Notari(2012{\natexlab{b}})]{masina2012b}
I.~Masina and A.~Notari.
\newblock Inflation from the higgs field false vacuum with hybrid potential.
\newblock \emph{arXiv preprint arXiv:1204.4155}, 2012{\natexlab{b}}.

\bibitem[Brennan et~al.(2017)Brennan, Carta, and Vafa]{brennan2017}
T.~D. Brennan, F.~Carta, and C.~Vafa.
\newblock The string landscape, the swampland, and the missing corner.
\newblock \emph{arXiv preprint arXiv:1711.00864}, 2017.

\bibitem[van Beest et~al.(2022)van Beest, Calder\'on-Infante, Mirfendereski, and Valenzuela]{vanBeest2022}
M.~van Beest, J.~Calder\'on-Infante, D.~Mirfendereski, and I.~Valenzuela.
\newblock Phys. rept. 989 (2022) 1–50.
\newblock \emph{arXiv}, 2022.

\bibitem[Agmon et~al.(2022)Agmon, Bedroya, Kang, and Vafa]{agmon2022}
N.~B. Agmon, A.~Bedroya, M.~J. Kang, and C.~Vafa.
\newblock Lectures on the string landscape and the swampland.
\newblock \emph{arXiv}, 2022.

\bibitem[Heisenberg et~al.(2018)Heisenberg, Bartelmann, Brandenberger, and Refregier]{heisenberg2018}
L.~Heisenberg, M.~Bartelmann, R.~Brandenberger, and A.~Refregier.
\newblock Dark energy in the swampland.
\newblock \emph{Physical Review D}, 98\penalty0 (12):\penalty0 123502, 2018.

\bibitem[Hertzberg et~al.(2019)Hertzberg, Sandora, and Trodden]{hertzberg2019}
M.~P. Hertzberg, M.~Sandora, and M.~Trodden.
\newblock Quantum fine-tuning in stringy quintessence models.
\newblock \emph{Physics Letters B}, 797:\penalty0 134878, 2019.

\bibitem[França and Rosenfeld(2002)]{franca2002}
U.~França and R.~Rosenfeld.
\newblock Fine tuning in quintessence models with exponential potentials.
\newblock \emph{Journal of High Energy Physics}, 2002\penalty0 (10):\penalty0 015, 2002.

\bibitem[Han et~al.(2019)Han, Pi, and Sasaki]{han2019}
C.~Han, S.~Pi, and M.~Sasaki.
\newblock Quintessence saves higgs instability.
\newblock \emph{Physics Letters B}, 791:\penalty0 314--318, 2019.

\bibitem[Axenides and Dimopoulos(2004)]{axenides2004}
M.~Axenides and K.~Dimopoulos.
\newblock Hybrid dark sector: Locked quintessence and dark matter.
\newblock \emph{Journal of Cosmology and Astroparticle Physics}, 2004\penalty0 (07):\penalty0 010, 2004.

\bibitem[Agrawal et~al.(2018)Agrawal, Obied, Steinhardt, and Vafa]{agrawal2018}
P.~Agrawal, G.~Obied, P.~J. Steinhardt, and C.~Vafa.
\newblock On the cosmological implications of the string swampland.
\newblock \emph{Phys. Lett. B}, 784:\penalty0 271, 2018.

\bibitem[Lee et~al.(2006)Lee, Lee, Lee, and Park]{Lee2006}
Wonwoo Lee, Bum-Hoon Lee, Chul~H. Lee, and Chanyong Park.
\newblock False vacuum bubble nucleation due to a nonminimally coupled scalar field.
\newblock \emph{Physical Review D}, 74:\penalty0 123520, 2006.
\newblock \doi{10.1103/PhysRevD.74.123520}.

\bibitem[Batini et~al.(2024)Batini, Chatrchyan, and Berges]{Batini2024}
Laura Batini, Aleksandr Chatrchyan, and Jürgen Berges.
\newblock Real-time dynamics of false vacuum decay.
\newblock \emph{Physical Review D}, 109:\penalty0 023502, 2024.
\newblock \doi{10.1103/PhysRevD.109.023502}.

\bibitem[Copeland et~al.(1994)Copeland, Liddle, Lyth, Stewart, and Wands]{copeland1994false}
Edmund~J Copeland, Andrew~R Liddle, David~H Lyth, Ewan~D Stewart, and David Wands.
\newblock False vacuum inflation with einstein gravity.
\newblock \emph{Physical Review D}, 49\penalty0 (12):\penalty0 6410, 1994.

\bibitem[Calmet(2018)]{calmet2018}
X.~Calmet.
\newblock Vanishing of quantum gravitational corrections to vacuum solutions of general relativity at second order in curvature.
\newblock \emph{Phys. Lett. B}, 787:\penalty0 36--38, 2018.
\newblock \doi{10.1016/j.physletb.2018.10.040}.

\bibitem[Kowalska et~al.(2022)Kowalska, Pramanick, and Sessolo]{kowalska2022naturally}
Kamila Kowalska, Soumita Pramanick, and Enrico~Maria Sessolo.
\newblock Naturally small yukawa couplings from trans-planckian asymptotic safety.
\newblock \emph{Journal of High Energy Physics}, 2022\penalty0 (8):\penalty0 1--30, 2022.

\bibitem[Gies et~al.(2017)Gies, Sondenheimer, and Warschinke]{gies2017impact}
Holger Gies, Ren{\'e} Sondenheimer, and Matthias Warschinke.
\newblock Impact of generalized yukawa interactions on the lower higgs-mass bound.
\newblock \emph{The European Physical Journal C}, 77:\penalty0 1--19, 2017.

\bibitem[Kohri and Matsui(2016)]{kohri2016higgs}
Kazunori Kohri and Hiroki Matsui.
\newblock Higgs vacuum metastability in primordial inflation, preheating, and reheating.
\newblock \emph{Physical Review D}, 94\penalty0 (10):\penalty0 103509, 2016.

\bibitem[Urbanowski(2022)]{urbanowski2022}
K.~Urbanowski.
\newblock Cosmological "constant" in a universe born in the metastable false vacuum state.
\newblock \emph{The European Physical Journal C}, 82\penalty0 (3):\penalty0 242, 2022.

\bibitem[Urbanowski(2023)]{urbanowski2023}
K.~Urbanowski.
\newblock A universe born in a metastable false vacuum state needs not die.
\newblock \emph{The European Physical Journal C}, 83\penalty0 (1):\penalty0 55, 2023.

\bibitem[Hollands and Wald(2015)]{hollands2015quantum}
Stefan Hollands and Robert~M Wald.
\newblock Quantum fields in curved spacetime.
\newblock \emph{Physics Reports}, 574:\penalty0 1--35, 2015.

\bibitem[Kay(1992)]{kay1992quantum}
Bernard~S Kay.
\newblock Quantum field theory in curved spacetime.
\newblock In \emph{Mathematical Physics X: Proceedings of the Xth Congress on Mathematical Physics, Held at Leipzig, Germany, 30 July--9 August, 1991}, pages 383--387. Springer, 1992.

\bibitem[Kiselev and Selivanov(1984)]{kiselev1984}
Valerij~G. Kiselev and Konstantin~G. Selivanov.
\newblock Calculation of the functional determinant in the vacuum-explosion problem.
\newblock 1984.

\bibitem[Devoto et~al.(2022)Devoto, Devoto, Luzio, and Ridolfi]{devoto2022}
F.~Devoto, S.~Devoto, L.~Di Luzio, and G.~Ridolfi.
\newblock False vacuum decay: an introductory review.
\newblock \emph{Journal of Physics G: Nuclear and Particle Physics}, 2022.

\bibitem[Degrassi and et~al.(2012)]{degrassi2012}
Giuseppe Degrassi and et~al.
\newblock Higgs mass and vacuum stability in the standard model at nnlo.
\newblock \emph{Journal of High Energy Physics}, 2012\penalty0 (8):\penalty0 1--33, 2012.

\bibitem[Peebles and Ratra(2003)]{peebles2003cosmological}
P~James~E Peebles and Bharat Ratra.
\newblock The cosmological constant and dark energy.
\newblock \emph{Reviews of modern physics}, 75\penalty0 (2):\penalty0 559, 2003.

\bibitem[Armendariz-Picon et~al.(2000)Armendariz-Picon, Mukhanov, and Steinhardt]{armendariz2000dynamical}
C.~Armendariz-Picon, V.~Mukhanov, and P.~J. Steinhardt.
\newblock A dynamical solution to the problem of a small cosmological constant and late-time cosmic acceleration.
\newblock \emph{Physical Review Letters}, 85\penalty0 (21):\penalty0 4438--4441, 2000.

\bibitem[Liddle and Lyth(2000)]{Liddle:2000cg}
Andrew~R. Liddle and D.~H. Lyth.
\newblock \emph{Cosmological inflation and large scale structure}.
\newblock Cambridge University Press, 2000.
\newblock \doi{10.1017/CBO9781139175180}.

\bibitem[Urbanowski(2017)]{urbanowski2017properties}
K~Urbanowski.
\newblock Properties of the false vacuum as a quantum unstable state.
\newblock \emph{Theoretical and Mathematical Physics}, 190\penalty0 (3):\penalty0 458--469, 2017.

\bibitem[Sivaram et~al.(2020)Sivaram, Rebecca, and Kenath]{sivaram2020a}
C.~Sivaram, L.~Rebecca, and A.~Kenath.
\newblock Planckian pre big bang phase of the universe.
\newblock \emph{Astrophysics and Space Science}, 365:\penalty0 17, 2020.

\bibitem[Myrzakulov et~al.(2015)Myrzakulov, Sebastiani, and Vagnozzi]{myrzakulov2015inflation}
Ratbay Myrzakulov, Lorenzo Sebastiani, and Sunny Vagnozzi.
\newblock Inflation in f (r, $\phi$) f (r, $\phi$)-theories and mimetic gravity scenario.
\newblock \emph{The European Physical Journal C}, 75:\penalty0 1--11, 2015.

\bibitem[Berezin et~al.(1988)Berezin, Kuzmin, and Tkachev]{berezin19883}
VA~Berezin, VA~Kuzmin, and II~Tkachev.
\newblock O (3)-invariant tunneling in general relativity.
\newblock \emph{Physics Letters B}, 207\penalty0 (4):\penalty0 397--403, 1988.

\bibitem[Senatore(2016)]{senatore2016lectures}
Leonardo Senatore.
\newblock Lectures on inflation.
\newblock \emph{Theoretical Advanced Study Institute in Elementary Particle Physics: new frontiers in fields and strings}, pages 447--543, 2016.

\bibitem[Struyve(2015)]{struyve2015}
Ward Struyve.
\newblock Semi-classical approximations based on bohmian mechanics.
\newblock \emph{International Journal of Modern Physics A}, 35, 07 2015.
\newblock \doi{10.1142/S0217751X20500700}.

\bibitem[Ooguri and Vafa(2007)]{ooguri2007geometry}
Hirosi Ooguri and Cumrun Vafa.
\newblock On the geometry of the string landscape and the swampland.
\newblock \emph{Nuclear physics B}, 766\penalty0 (1-3):\penalty0 21--33, 2007.

\bibitem[Szydłowski et~al.(2017)]{szydlowski2017}
Marek Szydłowski et~al.
\newblock Cosmological implications of quantum mechanics parametrization of dark energy.
\newblock \emph{J. Phys.: Conf. Ser.}, 880:\penalty0 012022, 2017.
\newblock \doi{10.1088/1742-6596/880/1/012022}.

\bibitem[Hashiba et~al.(2021)Hashiba, Yamada, and Yokoyama]{hashiba2021particle}
Soichiro Hashiba, Yusuke Yamada, and Jun’ichi Yokoyama.
\newblock Particle production induced by vacuum decay in real time dynamics.
\newblock \emph{Physical Review D}, 103\penalty0 (4):\penalty0 045006, 2021.

\bibitem[Sivaram et~al.(2011)Sivaram, Arun, and Nagaraja]{sivaram2011dieterici}
C.~Sivaram, K.~Arun, and R.~Nagaraja.
\newblock Dieterici gas as a unified model for dark matter and dark energy.
\newblock 335:\penalty0 599--602, 2011.
\newblock \doi{10.1007/s10509-011-0770-2}.

\bibitem[Albrecht and Steinhardt(1982)]{PhysRevLett.48.1220}
Andreas Albrecht and Paul~J. Steinhardt.
\newblock Cosmology for grand unified theories with radiatively induced symmetry breaking.
\newblock \emph{Phys. Rev. Lett.}, 48:\penalty0 1220--1223, 4 1982.
\newblock \doi{10.1103/PhysRevLett.48.1220}.
\newblock URL \url{https://link.aps.org/doi/10.1103/PhysRevLett.48.1220}.

\bibitem[Niedermann and Sloth(2024)]{niedermann2024new}
Florian Niedermann and Martin~S Sloth.
\newblock New early dark energy as a solution to the h 0 and s 8 tensions.
\newblock In \emph{The Hubble Constant Tension}, pages 431--456. Springer, 2024.

\bibitem[Niedermann and Sloth(2021)]{niedermann2021new}
Florian Niedermann and Martin~S Sloth.
\newblock New early dark energy.
\newblock \emph{Physical Review D}, 103\penalty0 (4):\penalty0 L041303, 2021.

\bibitem[Herold and Ferreira(2023)]{herold2023resolving}
Laura Herold and Elisa~GM Ferreira.
\newblock Resolving the hubble tension with early dark energy.
\newblock \emph{Physical Review D}, 108\penalty0 (4):\penalty0 043513, 2023.

\bibitem[Hill et~al.(2020)Hill, McDonough, Toomey, and Alexander]{hill2020early}
J~Colin Hill, Evan McDonough, Michael~W Toomey, and Stephon Alexander.
\newblock Early dark energy does not restore cosmological concordance.
\newblock \emph{Physical Review D}, 102\penalty0 (4):\penalty0 043507, 2020.

\bibitem[Kamionkowski and Riess(2023)]{kamionkowski2023hubble}
Marc Kamionkowski and Adam~G Riess.
\newblock The hubble tension and early dark energy.
\newblock \emph{Annual Review of Nuclear and Particle Science}, 73\penalty0 (1):\penalty0 153--180, 2023.

\bibitem[Sher(1989)]{Sher1989}
M.~Sher.
\newblock Electroweak higgs potentials and vacuum stability.
\newblock \emph{Physics Reports}, 179:\penalty0 273--418, 1989.
\newblock \doi{10.1016/0370-1573(89)90001-3}.

\bibitem[Arnold and Vokos(1991)]{Arnold1991}
P.~B. Arnold and S.~Vokos.
\newblock Instability of hot electroweak theory: bounds on $m(h)$ and $m(t)$.
\newblock \emph{Physical Review D}, 44\penalty0 (11):\penalty0 3620--3627, 1991.
\newblock \doi{10.1103/PhysRevD.44.3620}.

\bibitem[Espinosa and Quiros(1995)]{Espinosa1995}
J.~R. Espinosa and M.~Quiros.
\newblock Improved metastability bounds on the standard model higgs mass.
\newblock \emph{Physics Letters B}, 353:\penalty0 257--266, 1995.
\newblock \doi{10.1016/0370-2693(95)00693-6}.

\bibitem[Garny et~al.(2016)Garny, Sandora, and Sloth]{PhysRevLett.116.101302}
Mathias Garny, McCullen Sandora, and Martin~S. Sloth.
\newblock Planckian interacting massive particles as dark matter.
\newblock \emph{Phys. Rev. Lett.}, 116:\penalty0 101302, 3 2016.
\newblock \doi{10.1103/PhysRevLett.116.101302}.
\newblock URL \url{https://link.aps.org/doi/10.1103/PhysRevLett.116.101302}.

\bibitem[Di~Valentino et~al.(2017)Di~Valentino, Melchiorri, and Mena]{di2017can}
Eleonora Di~Valentino, Alessandro Melchiorri, and Olga Mena.
\newblock Can interacting dark energy solve the h 0 tension?
\newblock \emph{Physical Review D}, 96\penalty0 (4):\penalty0 043503, 2017.

\bibitem[Di~Valentino et~al.(2020{\natexlab{a}})Di~Valentino, Melchiorri, Mena, and Vagnozzi]{di2020interacting}
Eleonora Di~Valentino, Alessandro Melchiorri, Olga Mena, and Sunny Vagnozzi.
\newblock Interacting dark energy in the early 2020s: A promising solution to the h0 and cosmic shear tensions.
\newblock \emph{Physics of the Dark Universe}, 30:\penalty0 100666, 2020{\natexlab{a}}.

\bibitem[Di~Valentino et~al.(2020{\natexlab{b}})Di~Valentino, Melchiorri, Mena, and Vagnozzi]{di2020nonminimal}
Eleonora Di~Valentino, Alessandro Melchiorri, Olga Mena, and Sunny Vagnozzi.
\newblock Nonminimal dark sector physics and cosmological tensions.
\newblock \emph{Physical Review D}, 101\penalty0 (6):\penalty0 063502, 2020{\natexlab{b}}.

\bibitem[Baer et~al.(2015)Baer, Choi, Kim, and Roszkowski]{baer2015dark}
H.~Baer, K.~Y. Choi, J.~E. Kim, and L.~Roszkowski.
\newblock Dark matter production in the early universe: beyond the thermal wimp paradigm.
\newblock \emph{Physics Reports}, 555:\penalty0 1--60, 2015.

\bibitem[Wang and et~al.(2016)]{wang2016probing}
Y.~Wang and et~al.
\newblock Probing the interaction between dark energy and dark matter with the parameterized post-friedmann approach.
\newblock \emph{Physical Review D}, 94\penalty0 (8):\penalty0 083521, 2016.

\bibitem[Wang et~al.(2016)Wang, Abdalla, Atrio-Barandela, and Pav\'on]{wang2016dark}
B.~Wang, E.~Abdalla, F.~Atrio-Barandela, and D.~Pav\'on.
\newblock Dark matter and dark energy interactions: theoretical challenges, cosmological implications and observational signatures.
\newblock \emph{Rep. Prog. Phys.}, 79:\penalty0 096901, 2016.

\bibitem[Mukaida and Yamada(2017)]{muka2017Falsevaccum}
K.~Mukaida and M.~Yamada.
\newblock False vacuum decay catalyzed by black holes.
\newblock \emph{Physical Review D}, 96, 2017.
\newblock \doi{10.1103/physrevd.96.103514}.

\bibitem[Lepe and Pe{\~n}a(2016)]{lepe2016interacting}
Samuel Lepe and Francisco Pe{\~n}a.
\newblock Interacting cosmic fluids and phase transitions under a holographic modeling for dark energy.
\newblock \emph{The European Physical Journal C}, 76\penalty0 (9):\penalty0 507, 2016.

\bibitem[Banihashemi et~al.(2020)Banihashemi, Khosravi, and Shirazi]{banihashemi2020phase}
Abdolali Banihashemi, Nima Khosravi, and Amir~H Shirazi.
\newblock Phase transition in the dark sector as a proposal to lessen cosmological tensions.
\newblock \emph{Physical Review D}, 101\penalty0 (12):\penalty0 123521, 2020.

\bibitem[Brandenberger et~al.(2019)Brandenberger, Fr{\"o}hlich, and Namba]{brandenberger2019unified}
Robert Brandenberger, J{\"u}rg Fr{\"o}hlich, and Ryo Namba.
\newblock Unified dark matter, dark energy and baryogenesis via a “cosmological wetting transition”.
\newblock \emph{Journal of Cosmology and Astroparticle Physics}, 2019\penalty0 (09):\penalty0 069, 2019.

\bibitem[P{\'e}rez de~los Heros(2020)]{perez2020status}
Carlos P{\'e}rez de~los Heros.
\newblock Status, challenges and directions in indirect dark matter searches.
\newblock \emph{Symmetry}, 12\penalty0 (10):\penalty0 1648, 2020.

\bibitem[Klasen et~al.(2015)Klasen, Pohl, and Sigl]{klasen2015indirect}
Michael Klasen, Martin Pohl, and G{\"u}nter Sigl.
\newblock Indirect and direct search for dark matter.
\newblock \emph{Progress in Particle and Nuclear Physics}, 85:\penalty0 1--32, 2015.

\bibitem[Slatyer(2018)]{slatyer2018indirect}
Tracy~R Slatyer.
\newblock Indirect detection of dark matter.
\newblock \emph{Theoretical Advanced Study Institute in Elementary Particle Physics: anticipating the next discoveries in particle physics}, pages 297--353, 2018.

\end{thebibliography}

\section{Declaration of Generative AI and AI-assisted technologies in the writing process}
During the preparation of this work, the author, Pradosh Keshav, used Grammarly to correct grammar mistakes and make some paragraphs more structured. After using this tool/service, the author reviewed and edited the content as needed and take(s) full responsibility for the content of the publication.
\end{document}